\documentclass{aa}

\usepackage{booktabs}
\usepackage{graphicx}
\usepackage{lscape}
\usepackage{multirow}
\usepackage{xcolor}
\usepackage{graphicx}
\usepackage{ftnxtra}
\usepackage{fnpos}
\usepackage{txfonts}
\usepackage{soul}
\setstcolor{red}
\usepackage{supertabular}
\usepackage{tabularx}
\usepackage{siunitx}
\usepackage{collcell}
\usepackage[super]{nth}
\usepackage{placeins}
\usepackage{natbib,twoopt}
\usepackage[colorlinks, citecolor=blue, linkcolor=blue]{hyperref}
\makeatletter
\renewcommand*\aa@pageof{, page \thepage{} of \pageref*{LastPage}}
    \def\instrefs#1{{\def\scsep{\def\scsep{,}}\@for\w:=#1\do{\scsep\ref{inst:\w}}}}
    \renewcommand{\inst}[1]{\unskip$^{\instrefs{#1}}$}
\makeatother

\newcommand{\StarJOne}{G~261--6}
\newcommand{\PlanetJOne}{G~261--6\,b}
\newcommand{\MassJOne}{\SI{1.37\pm0.23}{M_\oplus}}
\newcommand{\PeriodJOne}{\SI{5.4536\pm0.0032}{\day}}
\newcommand{\ProtJOne}{\SI{114}{\day}}

\newcommand{\StarJTwo}{G~192--15}
\newcommand{\MassJTwoc}{\SI{14.3}{M_\oplus}}
\newcommand{\MassJTwob}{\SI{1.03\pm0.18}{M_\oplus}}

\newcommand{\StarJZero}{G~268--110}
\newcommand{\PlanetJZero}{G~268--110\,b}
\newcommand{\MassJZero}{\SI{1.52\pm0.25}{M_\oplus}}
\newcommand{\PeriodJZero}{\SI{1.43263\pm0.000076}{\day}}

\let\orgautoref\autoref
\renewcommand{\autoref}
        {\def\equationautorefname{Eq.}%
         \def\figureautorefname{Fig.}%
         \def\sectionautorefname{Sect.}%
         \def\subsectionautorefname{Sect.}%
         \def\subsubsectionautorefname{Sect.}%
         \orgautoref}
\begin{document} 
   \title{The CARMENES search for exoplanets around M dwarfs}
   \subtitle{Occurrence rates of Earth-like planets around very low-mass stars}
   \author{
A.~Kaminski\inst{lsw}
\and
S.~Sabotta\inst{lsw}
\and
J.~Kemmer\inst{lsw}
\and
P.~Chaturvedi\inst{tifr,tls}
\and
R.~Burn\inst{mpia}
\and
J.\,C.~Morales\inst{ice,ieec}
\and
J.\,A.~Caballero\inst{cabesac}
\and 
I.~Ribas\inst{ice,ieec}
\and
A.~Reiners\inst{iag}
\and
A.~Quirrenbach\inst{lsw}
\and
P.\,J.~Amado\inst{iaa}
\and
V.\,J.\,S.~B\'ejar\inst{iac,ull}
\and
S.~Dreizler\inst{iag}
\and
E.\,W.~Guenther\inst{tls}
\and
A.\,P.~Hatzes\inst{tls}
\and
Th.~Henning\inst{mpia}
\and
M.~K\"urster\inst{mpia}
\and
D.~Montes\inst{ucm}
\and
E.~Nagel\inst{iag}
\and
E.~Pall\'e\inst{iac,ull}
\and
V.~Pinter\inst{caha}
\and
S.~Reffert\inst{lsw}
\and
M.~Schlecker\inst{tucson-AS}
\and
Y.~Shan\inst{iag,oslo}
\and
T.~Trifonov\inst{lsw,sofia}
\and
M.\,R.~Zapatero Osorio\inst{cabesac}
\and
M.~Zechmeister\inst{iag}
}
   \authorrunning{A. Kaminski et al.}
   \titlerunning{Occurrence rates around low-mass M dwarfs}
   \institute{
            \label{inst:lsw}Landessternwarte, Zentrum f\"ur Astronomie der Universit\"at Heidelberg, K\"onigstuhl 12, 69117 Heidelberg, Germany \\
        \email{a.kaminski@lsw.uni-heidelberg.de}
        \and
            \label{inst:tifr}Department of Astronomy and Astrophysics, Tata Institute of Fundamental Research, Mumbai 400005, India
        \and
            \label{inst:tls}Th\"uringer Landessternwarte Tautenburg, Sternwarte 5, 07778 Tautenburg, Germany
        \and
            \label{inst:mpia}Max-Planck-Institut f\"ur Astronomie, K\"onigstuhl 17, 69117 Heidelberg, Germany   
        \and
            \label{inst:ice}Institut de Ci\`encies de l'Espai (CSIC), 
	c/ de Can Magrans s/n, Campus UAB, 
	08193 Bellaterra, Barcelona, Spain
         \and
            \label{inst:ieec}Institut d'Estudis Espacials de Catalunya, 
	08860 Castelldefels, Barcelona, Spain
        \and
            \label{inst:cabesac}Centro de Astrobiolog\'ia (CSIC-INTA), 
	Camino Bajo del Castillo s/n, 
	28692 Villanueva de la Ca\~nada, Madrid, Spain
        \and
            \label{inst:iag}Institut f\"ur Astrophysik und Geophysik, Georg-August-Universit\"at, Friedrich-Hund-Platz 1, 37077 G\"ottingen, Germany
        \and
            \label{inst:iaa}Instituto de Astrof\'isica de Andaluc\'ia (CSIC), 
	Glorieta de la Astronom\'ia s/n, 
	18008 Granada, Spain
        \and
            \label{inst:iac}Instituto de Astrof\'isica de Canarias, 
	38205 La Laguna, Tenerife, Spain
        \and 
            \label{inst:ull}Departamento de Astrof\'isica, Universidad de La Laguna, 
	38206 La Laguna, Tenerife, Spain
        \and
            \label{inst:ucm}Departamento de F\'isica de la Tierra y Astrof\'isica \& IPARCOS 
	Instituto de F\'isica de Part\'iculas y del Cosmos, Facultad de Ciencias
	F\'isicas, Universidad Complutense de Madrid, Plaza de Ciencias 1, 
	28400 Madrid, Spain
        \and
            \label{inst:caha}Centro Astron\'omico Hispano en Andaluc\'ia, 
	Observatorio Astron\'omico de Calar Alto, 
	Sierra de los Filabres, 04550 G\'ergal, Almer\'ia, Spain
        \and
            \label{inst:tucson-AS}Department of Astronomy/Steward Observatory, The University of Arizona, 933 North Cherry Avenue, Tucson, AZ 85721, United States of America
        \and
            \label{inst:oslo}Centre for Planetary Habitability, Department of Geosciences, Universitetet i Oslo, Sem S{\ae}lands vei 2b, 0315 Oslo, Norway
        \and
            \label{inst:sofia}Department of Astronomy, Sofia University ``St Kliment Ohridski'', 5 James Bourchier Blvd, 1164 Sofia, Bulgaria
         }

   \date{Received 10 December 2024 / Accepted 13 March 2025} 

  \abstract
  {}
   {Previous estimates of planet occurrence rates in the CARMENES survey indicated increased numbers of planets on short orbits for M dwarfs with masses below 0.34\,M$_\odot$. Here we focused on the lowest-mass stars in the survey, comprising 15 inactive targets with masses under 0.16\,M$_\odot$.}
   {To correct for detection biases, we determined detection sensitivity maps for individual targets and the entire sample. Using Monte Carlo simulations, we estimated planet occurrence rates for orbital periods of 1\,d to 100\,d and minimum masses from 0.5\,M$_\oplus$ to 10\,M$_\oplus$. We also compared the actual sample of known planets to model predictions.}
   {The radial velocity (RV) data from CARMENES reveal four new planets around three stars in our sample, namely G~268--110\,b, G~261--6\,b, and G~192--15\,b and c. All three b planets have minimum masses of 1.03--1.52\,M$_\oplus$ and orbital periods of 1.43--5.45\,d, while G~192--15\,c is a 14.3\,M$_\oplus$ planet on a wide, eccentric orbit with $P \approx 1218$\,d and $e \approx 0.68$.
   Our occurrence rates suggest considerable dependencies with respect to stellar masses. For planets below 3\,M$_\oplus$ we found rates consistent with one planet per star across all investigated periods, but the rates decrease almost by an order of magnitude for larger planet masses up to 10\,M$_\oplus$. Compared to previous studies, 
   low-mass stars tend to harbor more planets with $P <10$\,d. We also demonstrate that synthetic planet populations based on the standard core accretion scenario predict slightly more massive planets on wider orbits than observed.}
   {Our findings confirm that planet occurrence rates vary with stellar masses even among M dwarfs, as we found more planets with lower masses and on shorter orbits in our subsample of very low-mass stars compared to more massive M dwarfs. Therefore, we emphasize the need for additional differentiation in future studies.}   
    
   \maketitle

\section{Introduction} \label{sec:introduction}

The number of confirmed exoplanets has steadily increased since the first discoveries were made nearly three decades ago. Although during the initial era of that research most exoplanets were found by the means of Doppler spectroscopy, carried out in long time-baseline programs ~\citep[e.g.,][]{Vogt_1994,HARPS}, over recent years most of the discoveries have been made via space-based transit observations with missions such as CoRoT \citep{2006cosp...36.3749B}, \textit{Kepler} \citep{Borucki2010}, and \textit{TESS} \citep{Ricker2015}.

The observed population of currently almost 6000 confirmed exoplanets and candidates serves as a probe for theoretical models on planet formation. Ideally, the synthetic planet populations from models should converge towards the observed one. Unfortunately, the task is not as trivial as it sounds. In order to reproduce the actual planet population by modeling, one needs to understand the physical process in its entirety. In particular, this also includes differentiation on dynamic time scales and stellar, or rather protoplanetary, disk masses. On the other hand, the outcomes of observational surveys heavily depend on the used methods, instruments' sensitivity, target selection, as well as number and methodology of observations. Those choices of course lead to some selection biases, which need to be accounted for when determining the planet populations from them. 

Naturally, the confidence in conclusions derived from observational population studies is highly affected by the number of considered objects. Therefore, the discovery and characterization of thousands of transiting planet candidates by \textit{Kepler} was a crucial step forward. Since the mission focused on G-type stars, \textit{Kepler} transiting planets around M dwarfs were highly underrepresented, and thus occurrence rates of planets orbiting M dwarfs relying on that survey come with high uncertainties \citep{Hardegree-Ullman2019}. Still, it was shown that the occurrence rate of small planets on short orbits up to 50\,d around M dwarfs is higher than around solar-like stars \citep{Howard2012}. 

Across the whole spectrum of stellar types, M dwarfs are particularly interesting to study. They are not only the most common type of stars \cite[and references therein]{Reyle2021A&A...650A.201R}, but also favorable, because due to their low mass and size, small and low-mass planets can be detected around them more easily using Doppler spectroscopy and the transit method, respectively. It was found very early on, and has been confirmed thereafter, that the planet population around M dwarfs differs from those around other types of host stars. As the occurrence rate of giant planets was observed to be correlated with stellar mass \citep{Johnson_2010}, in particular hot Jupiters 
were believed to be rare around M dwarfs compared to their occurrence rate around hotter and more massive stars \citep{Endl2006, Johnson_2012, Hartman_2015}. While the actual percentage of cool stars hosting hot Jupiters is still under debate due to high uncertainty levels in occurrence rate analyses, as determined by, for example, \cite{Obermeier_2016}, it appears that the frequency of hot Jupiters peaks at G-type dwarfs and decreases in both directions, towards hotter and cooler stars \citep{Gan_2023}.

Differential studies of planet occurrence across the full M-dwarf spectral type are particularly valuable because they span a wide range of masses, temperatures, and stellar environments, all of which can significantly influence planet formation and evolution. By examining the occurrence rates of planets around early-, mid-, and late-type M dwarfs, one can explore how factors such as stellar mass, disk properties, and stellar activity shape the frequency and characteristics of formed planets. 
These comparative studies help to identify important trends, such as whether low-mass planets are more prevalent around specific subtypes of M dwarfs and how stellar properties influence the potential habitability of orbiting planets. Such analyses do not only refine planet formation theories, but also improve our understanding of where potentially habitable planets are most likely to be found.
Estimations of occurrence rates of planets around M dwarfs have only been reported for the earlier spectral types, of about M3.5--4.0\,V, which corresponds to stellar masses down to around 0.33\,M$_\odot$ \citep[e.g.,][]{Bonfils_2013_sample, Pinamonti_2022}. For M dwarfs of later spectral types, however, the picture becomes comparatively incomplete. Because of their faintness, it is more challenging to search for planets around late, low-mass M dwarfs.

Still, among M dwarfs, very low-mass stars ($M \lesssim 0.16$\,M$_\odot$) are of particular interest. Despite their smaller numbers in brightness-limited samples, their lower masses and cooler temperatures create unique conditions for planet formation, offering crucial insights into planetary systems that form in the least massive and faintest stellar environments. Understanding planet formation around these stars is essential for developing a complete picture of how planets form across the full range of stellar masses. Moreover, the low luminosity of very low-mass stars shifts the habitable zone to much closer orbits,
facilitating not only the RV detection of Earth-like planets, but also demographic studies on climatic conditions within and outside of the habitable zone~\citep[e.g.,][]{Checlair2019,Turbet2019a,Schlecker2024}.
These planets also orbit in environments where stellar activity and flaring may influence atmospheric retention and habitability \citep{Tarter2007}. Additionally, very low-mass stars provide advantageous targets for atmospheric characterization. The large planet-to-star size ratios and the proximity of habitable zones result in deeper transit signals and stronger atmospheric features in transmission spectroscopy, facilitating the study of the atmospheres of small planets, including those that could potentially harbor life \citep{Trifonov2021, Kuzuhara2024}.

The CARMENES\footnote{Calar Alto high-Resolution search for M dwarfs with Exoearths with Near-infrared and optical Échelle Spectrographs; \url{https://carmenes.caha.es}} spectrograph, installed at the 3.5\,m telescope of the Calar Alto Observatory in Almer\'ia, Spain, is ideally suited for investigating planetary systems around M dwarfs, including very low-mass stars. Specifically designed to conduct an RV survey of around 350 M dwarfs, the instrument has been operational since January 2016, covering both the visual (VIS) and near-infrared (NIR) wavelength ranges between 520\,nm and 1710\,nm with spectral resolutions of $R=94\,600$ in the VIS and $R=80\,400$ in the NIR \citep{Quirrenbach2014}. Soon after the first planet discovery findings, \cite{Sabotta2021} conducted a foundational study based on a sample of 71 M dwarfs observed with CARMENES, and reported an abundance of short-period planets, particularly around the latest-type M dwarfs.
They highlighted the need for further investigation into how occurrence rates vary across the M dwarf spectral type.
Based on these results, \cite{Ribas2023} expanded this analysis using a larger sample of 238 M dwarfs from the CARMENES guaranteed time observations (GTO). Their refined study confirmed the overabundance of short-period planets around late-type M dwarfs, further emphasizing the importance of differentiating planet occurrence rates by stellar mass. With this refinement, \cite{Ribas2023} provided an overall occurrence rate of
$1.44\pm0.20$ planets per star, illustrating that nearly every M dwarf hosts at least one planet, while also revealing significant trends related to stellar mass.

Our present study serves as a specific follow-up to the studies by both \cite{Sabotta2021} and \cite{Ribas2023}, focusing exclusively on the least massive stars ($M \lesssim 0.16$\,M$_\odot$, spectral type about M5.5\,V and later) from the CARMENES survey. 
With the focus on Earth-like planets that are detectable by current RV surveys, we studied only companions with orbital periods of up to 100\,d and planetary masses below 10\,M$_\oplus$. 
By refining the occurrence rates of Earth-like planets in this low-mass regime, we took an important first step toward completing the full picture of planetary distribution around M dwarfs. Although the current stellar sample is small and subject to statistical uncertainties, this work provides a primary basis for a more comprehensive analysis when the full CARMENES survey is complete. Our ultimate goal is to provide refined occurrence rates of Earth-like planets specifically for very low-mass stars, contributing to a broader understanding of planetary system formation across the entire M-dwarf spectrum.

The selection criteria for the targets of this study, as well as the final stellar sample itself, are summarized in \autoref{sec:sample}. Four newly discovered exoplanets around three stars of the sample are presented in \autoref{sec:meth_and_res}.
In \autoref{subsubsec: detcompleteness} the methods and steps for estimating the occurrence rates are documented and explained. Thereafter, in \autoref{sec: discussion}, the results are discussed with respect to their uncertainties and robustness of the methods, and are put into context by comparing them to previous analyses on the subject. In addition to that, the distribution of detected planets around the targets in our stellar sample are compared to predictions from planet formation theory.
Finally, we summarize and conclude our work in \autoref{sec:conclusions}.

\section{Stellar sample}\label{sec:sample}

\begin{table*}
\caption{Main properties of our sample stars.}
\label{table:stellar_properties}   
\centering 
\begin{tabular}{
  @{}ll@{}ccccc
  >{\collectcell\num}r<{\endcollectcell}
  @{${}\pm{}$}
  >{\collectcell\num}l<{\endcollectcell}
  l@{}cc@{}
}

\hline\hline                
\noalign{\smallskip}
Karmn & Star name & GJ &$\alpha$ (J2000) & $\delta$ (J2000) & Sp. type & $M_\star$ & \multicolumn{2}{c}{$P_\text{rot}$}& $P_\text{rot,ref}$ & pEW(H$\alpha$) & $N_\text{RVs}$ \\
 & & & & & & [M$_\odot$] & \multicolumn{2}{c}{[d]} & & [\AA] & \\
\noalign{\smallskip}
\hline                                   
\noalign{\smallskip}
J00067--075 & \object{G~158--27}& 1002           & 00:06:43.20 &--07:32:17.0 &M5.5\,V & $0.105\pm0.009$  & 93.0&1.7  & Fou23 &   --0.07 & 89   \\
J00184+440 & \object{G~171--48} & 15~B         & 00:18:25.82 &
+44:01:38.1 &M3.5\,V & $0.161\pm0.010$  & 113.3&4.3  & Don23 &  +0.15 &  193 \\
J01048--181 & \object{G~268--110}& 1028         & 01:04:53.80 &--18:07:28.6 &M5.0\,V & $0.137\pm0.009$  & 143&14       & New18     &  +0.006 & 113  \\ 
J01125--169 & \object{YZ~Cet} & 54.1            &01:12:30.64 & --16:59:56.4 &M4.5\,V & $0.138\pm0.009$  & 70.1&7.0  & Sha24 &  --1.40 & 110 \\
J02530+168 & \object{Teegarden's Star} & ... & 02:53:00.89 &+16:52:52.6 &M7.0\,V & $0.097\pm0.010$  & 97.6&9.8  & Laf21 &  --0.52  & 316 \\
J03133+047 & \object{CD~Cet} & 1057            & 03:13:22.92 &+04:46:29.3 &M5.0\,V & $0.161\pm0.009$  & 126&13        & New16     &  --0.02 & 107   \\ 
J06024+498 & \object{G~192--15} & 3380         & 06:02:29.19 &+49:51:56.2 &M5.0\,V & $0.132\pm0.009$  & 105&6    & DA19      & --0.007  & 147 \\
J06594+193 & \object{G~109--35} & 1093        & 06:59:28.82 &+19:20:55.9 &M5.0\,V & $0.118\pm0.009$  & 110&16     & This work & --0.32  & 28 \\
J08413+594 & \object{G~234--45} & 3512        & 08:41:20.13 &+59:29:50.4 &M5.5\,V & $0.123\pm0.009$  & 83.2&8.3     & Pas23 & --1.34  & 223 \\ 
J18027+375 & \object{G~182--36} & 1223        & 18:02:46.26 &+37:31:03.0 &M5.0\,V & $0.145\pm0.009$  & 124&12        & New16     & +0.05  & 118 \\ 
J19242+755 & \object{G~261--6} & 1238       & 19:24:16.31 &+75:33:11.8 &M5.5\,V & $0.118\pm0.011$  & 114&34          & Irw11     & --0.28  & 217 \\
J20260+585 & \object{Wolf~1069} & 1253       & 20:26:05.30 &+58:34:22.7 &M5.0\,V & $0.160\pm0.010$  & 160&16    & Med22 & --0.08  & 268 \\
J20556--140S & \object{LP~756--18} & 810~B      & 20:55:37.12 &--14:03:54.9 &M5.0\,V & $0.148\pm0.008$  & 134&13    & New18 & +0.08  & 53 \\ 
J23351--023 & \object{G~157--77} & 1286         & 23:35:10.46 &--02:23:20.6 &M5.5\,V & $0.114\pm0.009$  & 178&15   & Don23 & --0.72 & 71  \\
J23419+441 & \object{Ross~248} & 905        & 23:41:55.04 &+44:10:38.8 &M5.0\,V & $0.144\pm0.009$  & 106&6   & DA19      & --0.45 & 99  \\
\noalign{\smallskip}
\hline
\end{tabular}
\tablefoot{
DA19: \cite{DiezAlonso2019}; 
Don23: \cite{Donati2023};  
Fou23: \cite{Fouque2023}; 
Irw11: \cite{Irwin2011}; 
Laf21: \cite{Lafarga2021}; 
Med22: \cite{Medina2022};
New16: \cite{Newton2016}; 
New18:  \cite{Newton2018}; 
Pas23: \cite{Pass2023};
Sha24: \cite{Shan2024}.
The uncertainties for the stellar rotation periods were taken from literature. When absent, we imposed an uncertainty of \SI{10}{\percent} of the value of $P_\text{rot}$, as justified in detail by \cite{Shan2024}.
Coordinates are listed as provided by \cite{Gaia2021A&A...649A...6G}}.

\end{table*}

For this study, our aim was to intensively observe stars for which we could detect Earth-mass planets. For this reason, we set several constraints on our sample. As a baseline catalog, we took the CARMENES input catalog, namely Carmencita  \citep[version 106;][]{Carmencita}. 
Carmencita contains several dozens parameters for about 2200 nearby, bright M dwarfs, from which the GTO targets were selected. An updated summary of Carmencita was provided by \cite{CortesContreras2024}.
From there, we selected all targets with masses $M \le 0.1617$\,M$_\odot$. 
We chose this exact value because an Earth-mass planet in an orbit of 10\,d around a star with this stellar mass would induce an RV amplitude of 1\,m\,s$^{-1}$, which, based on our experience, is the minimum amplitude that can be detected by CARMENES with a reasonable number of observations \citep{Zechmeister2019,Luque2022_pl, Kemmer2022,Kossakowski2023,Mascareno23}. 
Stellar masses in Carmencita are computed following the methodology of \cite{Schweitzer2019}, which is determining bolometric luminosities from the integration of the spectral energy distribution and precise \textit{Gaia} parallaxes \citep{Cifuentes2020A&A...642A.115C}, effective temperatures from spectral synthesis \citep[e.g.,][and references therein]{Passegger2022}, stellar radii from the two parameters above and Stefan-Boltzmann law, and stellar masses from a mass-radius relationship calibrated with double-lined, detached, eclipsing binaries. 

Additionally, we applied a brightness threshold of $J \le$ 10\,mag,
because for fainter stars a precision of 1\,m\,s$^{-1}$ typically cannot be achieved unless an extraordinarily large number of RVs is collected. We also included only inactive (or, rather, very weakly active) targets, as activity can reduce the detection efficiency, and it is moreover difficult to characterize proper detection sensitivities for those targets. We set three constraints on the activity and kept stars with pseudo-equivalent of the H$\alpha$ line pEW(H$\alpha$) $>$ --1.5\,\AA, 
which removed around half of the targets within the mass limit, $P_\text{rot} > 10$\,d, and $v\sin{i} < 2$\,km\,s$^{-1}$. These thresholds made sure that we were not including young active stars with high chromospheric emission, which are expected to also be fast rotators. Less conservative limits on the three parameters were applied to studies of M dwarfs in the CARMENES sample by \cite{Schoefer2019}, \cite{ CortesContreras2024}, and \cite{Kemmer2025}.

The 15 remaining targets defined our sample and are listed in \autoref{table:stellar_properties}, together with their fundamental parameters and number of valid RVs collected by CARMENES.
The stellar rotation periods were collected from the literature, as referenced in the Carmencita catalog ($P_{\rm rot, ref}$ column in Table~\ref{table:stellar_properties}). They were used in the following to identify activity-induced RV signals and to distinguish them from planetary companions in our analysis. Only for one of our targets, G~109--35 (Karmn J06594+193), the rotation period was not known. Using available photometric data and spectroscopic activity indicators, we determined the missing rotation periods and arrived at $P_\text{rot}$ =110$^{+16}_{-13}$\,d. This analysis is presented in \autoref{app:rot_period}.
One more star with spectral type M5.0\,V, namely GJ~3250 (Karmn J03473+086), also met our defined requirements but was left out of the subsequent analysis, as it had been observed infrequently with CARMENES and only 12 RV measurements were available at the time of the analysis.

\section{Planet discoveries}\label{sec:meth_and_res}
\subsection{RV data} \label{sec:obs}

The RV measurements used for this work were collected from the M-dwarf survey carried out with the CARMENES spectrograph.
It was specifically designed to deliver highly accurate RVs with a precision of the order of 1\,m\,s$^{-1}$ to search for temperate rocky planets around nearby cool stars \citep{Ribas2023}. Although the instrument provides RVs from two separate spectral channels, we utilized only data from the VIS channel, covering the spectral region up to 960\,nm. All acquired spectra were reduced by the extraction pipeline \texttt{caracal} (CARMENES Reduction And CALibration; \citealt{Caballero2016}), and the RVs were determined by means of template matching with \texttt{serval} (SpEctrum Radial Velocity AnaLyser; \citealt{Zechmeister2018}). In order to correct for uncalibrated systematic effects, shared by RVs from the same night, nightly zero-point offsets (NZPs), as described by, for example, \cite{Trifonov2018A&A...609A.117T} and \cite{TalOr2019MNRAS.484L...8T}, were computed and applied. All observations of our stellar targets that were used for this publication were obtained between 12 January 2016 and 6 February 2024.

\subsection{Discovery of G~268--110\,b}\label{subsec:G268-110}

\begin{figure*}
    \centering
    \includegraphics[width=0.93\textwidth]{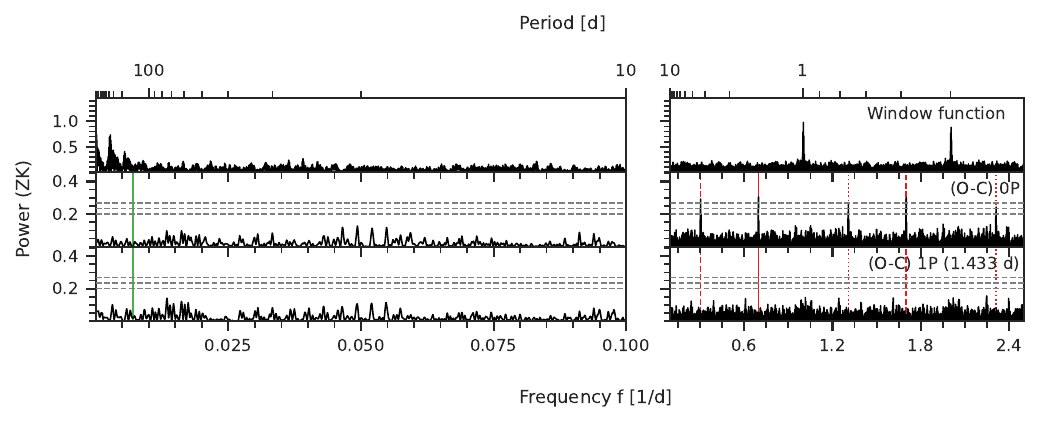} 
    \caption{Window function (upper panel) and GLS periodograms of the CARMENES RVs for \StarJZero{} before ('0P', middle) and after subtracting the one-planet model ('1P', bottom). 
    The two panels on the left and right represent the same GLS periodograms but plot different regions to better represent the occurring signals. The period of the 1.43-day planet is highlighted by a red solid line. Its first-order aliases at \SI{3.28}{\day} and \SI{0.59}{\day} are marked by red dashed lines, and the second-order aliases at \SI{0.76}{\day} and \SI{0.43}{\day} by red dotted lines. The rotation period of \SI{143}{\day} determined by \cite{Newton2018} is indicated by a solid green line.}
    \label{fig:gls_rv_J01048-181}
\end{figure*}

From six years of CARMENES observations, we report the discovery of the Earth-mass planet \PlanetJZero{} (GJ~1028, J01048--181). We first derived a tentative orbital period of $P=\PeriodJZero$ and a minimum planetary mass of $M_{\rm pl}\sin i =\MassJZero$. However, the RV data suffer from strong aliasing, as is evident in the Generalized Lomb-Scargle (GLS) periodogram \citep{Zechmeister2009} depicted in \autoref{fig:gls_rv_J01048-181}.
It shows a set of very significant signals with low false alarm probabilities (FAPs) at periods of \SI{1.43}{\day} (FAP $< \num{6.1E-05} $), \SI{0.59}{\day} (FAP $< \num{7.8E-05}$), and \SI{3.28}{\day} (FAP $< \num{1.6E-04}$), which are related by aliasing due to a sampling frequency of \SI{\sim1}{\per\day} that is dominant in our RV data. This aliasing is further evident through the also significant second-order aliases (with respect to the 1.43-day signal) at periods of \SI{0.76}{\day} and \SI{0.43}{\day}. After subtracting the 1.43-day signal, which has the lowest FAP of all, there are no other significant signals present.

Unfortunately, the signals are all of comparable low FAP in the GLS periodogram, which makes the determination of the true orbital period very difficult. We applied the \texttt{AliasFinder} code, which allows comparing synthetically generated periodograms for different alias periods with the observed one \citep{Dawson2010, Stock2020, Stock2020a}. The best match can be an indication of the likely underlying period of the signal. However, as can be seen in \autoref{fig:aliasing_J01048-181}, the periodograms resulting from the different first-order alias periods remain indistinguishable in our case. 
Therefore, we performed a model comparison for the three different periods, where we also allowed for non-zero eccentricity in the planetary orbit. We tested all of them against the base model (= ``0P''), which only includes a jitter and an offset of the CARMENES data. 

We used \texttt{juliet} \citep{juliet} to perform the fits to the RV data. Thus, Keplerian orbits were parameterized by their period $P$, RV semi-amplitude $K$, time of inferior conjunction $t_0$, and the $\mathcal{S}_1 = \sqrt{e}\sin\omega$ and $\mathcal{S}_2 = \sqrt{e}\cos\omega$ parameters, which depend on the eccentricity $e$ and argument of periastron $\omega$. For each of the three tested periods we tried both circular and eccentric orbits and constrained the priors for each signal to a narrow range around the highest local peak in order to also avoid issues from yearly aliasing, which is visible in \autoref{fig:aliasing_J01048-181}. An overview of the priors used is given in \autoref{tab:planetparams_priors}.
The results of this test are tabulated in \autoref{tab:modelcomp_G268-110}. 

\begin{table}
\caption{Model comparison for \StarJZero{}.} 
\label{tab:modelcomp_G268-110}
\centering
\begin{tabular}{l S[table-format=-3.1] S[table-format=-2.1] S[table-format=-3.1]}
\hline\hline
    \noalign{\smallskip}
               Model &  {$\ln\mathcal{Z}$} &  {$\Delta\ln\mathcal{Z}$} &  {max($\ln\mathcal{L}$)} \\
    \noalign{\smallskip}
\hline
    \noalign{\smallskip}
0P &            -332.9 &                -12.7 &        -327.5 \\
1P$_\text{(1.43 d-circ)}$ &            -320.2 &                 0 &        -307.9 \\
 1P$_\text{(1.43 d-ecc)}$ &            -320.6 &                -0.4 &        -304.0 \\
\noalign{\medskip}
1P$_\text{(0.59 d-circ)}$ &            -322.3 &                -2.1 &        -308.7 \\
   1P$_\text{(0.59 d-ecc)}$ &            -322.2 &                -2.0 &        -304.2 \\
\noalign{\medskip}
1P$_\text{(3.28 d-circ)}$ &            -320.9 &                -0.7 &        -308.5 \\
   1P$_\text{(3.28 d-ecc)}$ &            -320.3 &                -0.1 &        -303.2 \\
    \noalign{\smallskip}
\hline
\end{tabular}
\end{table}

We found that irrespective of the chosen period, all 1P-models are favored over the base model \citep[$\Delta\ln\mathcal{Z}>5$;][]{Trotta2008}, while eccentric models are indistinguishable ($\Delta\ln\mathcal{Z}<3$) from circular models. We therefore assumed a circular orbit because $e$ is not well determined by our data and thus could lead to an overestimation of its value \citep[][and references therein]{Hara2019}.

Regarding the different possible periods, $\ln\mathcal{Z}$ is only a valid metric for models with the same period priors. Therefore, for the comparison between the different periods we used the maximum log-likelihood ($\ln\mathcal{L}$) for the circular or eccentric models individually. 
However, it turned out that there are no significant differences between the individual maximum likelihoods. In combination with the alias test, this means that it is not possible to identify the true period of the signal from the current data set. A concentrated campaign with high-cadence observations, as would be necessary for the resolution of the 0.5-day alias, is unfortunately difficult to implement with CARMENES due to the low declination of \StarJZero. 

\begin{figure}
\centering
\includegraphics[width=0.44\textwidth]{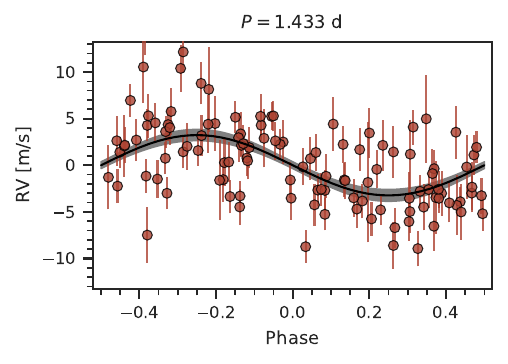}
\caption{Phased RV plot for \PlanetJZero{} based on the best fit model (1P$_\text{(1.43\,d-circ)}$). The black line depicts the model based on the parameters listed in \autoref{tab:planetparams}. The shaded area illustrates the 1$\sigma$ confidence interval.}  
\label{fig:phasefolded_J01048-181}
\end{figure}

Still, since the 1.43-day signal shows the lowest FAP in the GLS and slightly outperforms the models to the other periods in terms of $\ln\mathcal{L}$, we considered that period as the most probable one. The corresponding phase-folded RVs are plotted in \autoref{fig:phasefolded_J01048-181}, the complete RV time series can be found in \autoref{fig:rvs_long_multipanel}, and the model parameters, as well as the derived planetary parameters, are listed in \autoref{tab:planetparams}. However, for the sake of completeness, the parameters determined from the sampling of the alternative periods are listed in \autoref{tab:alt_planetparameters_J01048-181}.

\begin{table*}
\caption{Posterior parameters for the new planet discoveries.} 
\label{tab:planetparams}
\centering
{\setlength{\extrarowheight}{4.0pt}
\begin{tabular}{lccccl}
\hline \hline
Parameter & G~268--110\,b & G~261--6\,b & G~192--15\,b & G~192--15\,c & Description \\
\hline
\noalign{\smallskip}
\multicolumn{5}{c}{\textit{Posterior parameters}} \\
\noalign{\smallskip}
$P$ [d] & $\num{1.432630}^{+\num{72e-06}}_{-\num{76e-06}}$ & $\num{5.4536}^{+\num{0.0031}}_{-\num{0.0032}}$ & $\num{2.27476}^{+\num{26e-05}}_{-\num{28e-05}}$ & 
$1219^{+\num{13}}_{-\num{11}}$ & Orbital period \\
$t_{0}$ [BJD]& $\num{2457613.435}^{+\num{0.094}}_{-\num{0.087}}$ & $\num{2459345.14}^{+\num{0.22}}_{-\num{0.22}}$ & $\num{2457851.82}^{+\num{0.23}}_{-\num{0.20}}$ & 
$\num{2458813}^{+\num{10}}_{-\num{12}}$ & Time of potential transit-center  \\
$K$ [$\mathrm{m\,s^{-1}}$]& $\num{3.23}^{+\num{0.50}}_{-\num{0.50}}$ & $\num{2.08}^{+\num{0.31}}_{-\num{0.32}}$ & $\num{1.94}^{+\num{0.32}}_{-\num{0.32}}$ & $\num{4.49}^{+\num{0.61}}_{-\num{0.54}}$ & RV semi-amplitude  \\
$e$ & 0  & 0  & 0  & $\num{0.676}^{+\num{0.063}}_{-\num{0.073}}$ & Eccentricity\tablefootmark{({a})} \\
$\omega$ [deg] & ... & ... & ... & $\num{73.4}^{+\num{9.8}}_{-\num{9.6}}$ & Argument of periastron\tablefootmark{({a})} \\
\noalign{\smallskip}
\multicolumn{5}{c}{\textit{Derived parameters}} \\
\noalign{\smallskip}
$M_{\rm pl}\sin i$ [M$_\oplus$]& $\num{1.52}^{+\num{0.25}}_{-\num{0.25}}$ & $\num{1.37}^{+\num{0.23}}_{-\num{0.22}}$& $\num{1.03}^{+\num{0.18}}_{-\num{0.18}}$ & $\num{14.3}^{+\num{1.6}}_{-\num{1.5}}$ & Minimum mass \\
$a_{\rm p}$ [\si{\astronomicalunit}]& $\num{0.01283}^{+\num{0.00028}}_{-\num{0.00029}}$ & $\num{0.02971}^{+\num{0.00087}}_{-\num{0.00093}}$ & $\num{0.01723}^{+\num{0.00038}}_{-\num{0.00039}}$ & $\num{1.137}^{+\num{0.026}}_{-\num{0.027}}$ & Semi-major axis \\
$S$ [S$_\oplus$]& $\num{13.57}^{+\num{0.64}}_{-\num{0.57}}$ & $\num{1.80}^{+\num{0.12}}_{-\num{0.11}}$ & $\num{7.06}^{+\num{0.34}}_{-\num{0.31}}$ & $\num{0.001620}^{+\num{80e-06}}_{-\num{71e-06}}$ & Stellar irradiance \\
$T_\textnormal{eq, p}$ [\si{\kelvin}]& 
$\num{534}^{+\num{12}}_{-\num{11}}$ & $\num{322.3}^{+\num{9.9}}_{-\num{9.7}}$ & $\num{453.6}^{+\num{8.8}}_{-\num{8.5}}$ & $\num{55.9}^{+\num{1.1}}_{-\num{1.1}}$ & Equilibrium temperature\tablefootmark{({b})} \\
\noalign{\smallskip}
\multicolumn{5}{c}{\textit{Instrument parameters}} \\
\noalign{\smallskip}
$\gamma$ [$\mathrm{m\,s^{-1}}$]       &  +$\num{0.04}^{+\num{0.36}}_{-\num{0.35}}$  & +$\num{0.14}^{+\num{0.23}}_{-\num{0.23}}$           & \multicolumn{2}{c}{$\num{-0.20}^{+\num{0.23}}_{-\num{0.23}}$}      & Instrumental zero point      \\
$\sigma$ [$\mathrm{m\,s^{-1}}$]&   $\num{2.96}^{+\num{0.33}}_{-\num{0.30}}$     & $\num{1.60}^{+\num{0.34}}_{-\num{0.36}}$  &  \multicolumn{2}{c}{$1.62^{+\num{0.26}}_{-\num{0.27}}$} & RV jitter term \\
\noalign{\smallskip}
\hline
\end{tabular}}
\tablefoot{Error bars denote the $68\%$ posterior credibility intervals.
\tablefoottext{a}{Eccentricity fixed to null for G~268--110\,b, G~261--6\,b, and G~192--15\,b. 
As a result, their arguments of periastron are not defined.}
\tablefoottext{b}{Equilibrium temperature assuming a zero Bond albedo, $A_B=0$.}
}
\end{table*}

For the rotation period, we relied on an earlier measurement of \SI{143}{\day} by \cite{Newton2018} from photometric data from the MEarth project \citep{Irwin2009}. Although this is far from the signals detected in the RVs, we nevertheless explored the CARMENES activity indicators to exclude other forms of activity that are not related to the stellar rotation as the origin of the RV signals.
The list of activity indicators studied included all of those regularly computed in the CARMENES data flow, such as the chromatic index (CRX), differential linewidth (dLW), CCF bisector (BIS), and contrast (CON), as well as indicators related to the pEWs of specific lines such as H$\alpha$, TiO, and more \citep[see][for the full list]{Zechmeister2018, Schoefer2019, Lafarga2020A&A...636A..36L,Lafarga2021}.
In order to identify common periodicities appearing in the set of indicators, we first scanned the GLS periodograms of the activity indicators for common periods applying the {\tt DBSCAN} clustering algorithm \citep[Density-Based Spatial Clustering of Applications with Noise;][]{Ester1996} implemented in \texttt{scikit-learn} \citep{Pedregosa2012}, as described by \cite{Kemmer2025}. For this, we calculated the GLS periodogram for each activity indicator and identified the ten highest peaks.
The periodograms start at a period of \SI{2}{\day} to avoid aliases that occur due to the typical sampling of daily observations. The {\tt DBSCAN} algorithm was then run on the combined sample of all peaks with FAPs below $\SI{80}{\percent}$ and periods shorter than the baseline of the observations. In the process, we considered a minimum number of three samples in a neighborhood to be a cluster if the maximum frequency distance between two detected peaks is less than half of the width of the peaks in the GLS (i.e., $\delta f = (t_\text{max}-t_\text{min})^{-1}$).
This clustering analysis is illustrated in \autoref{fig:activity_clusters_J01048-181}. We did not detect any significant clusters of periods that would hint at a strong influence from stellar activity onto our data. All found clusters are related to the harmonics of one year or the Moon cycle, and none overlap with the planetary signal or its aliases in the RVs. This is also reflected in the GLS periodograms of the activity indicators with signals less than \SI{10}{\percent} FAP (see \autoref{fig:activity_GLS_J01048-181_onecol}). We therefore reasonably assumed that stellar activity does not need to be considered in our modeling.

\subsection{Discovery of G~261--6\,b}

\begin{figure}
\centering
\includegraphics[width=0.44\textwidth]{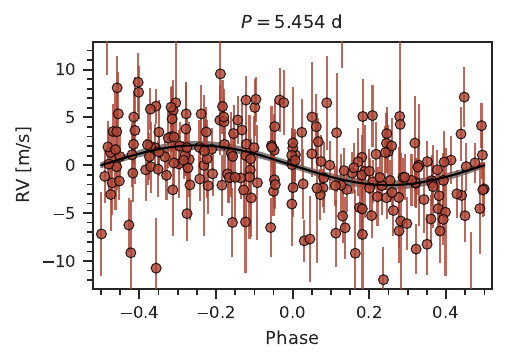}
\caption{Same as Fig.~\ref{fig:phasefolded_J01048-181} but for \PlanetJOne{} based on the best fit model (1P$_\text{(5.45\,d-circ)}$).} 
\label{fig:phasefolded_J19242+755}
\end{figure}

\begin{figure*}
    \centering
    \includegraphics[width=0.93\textwidth]{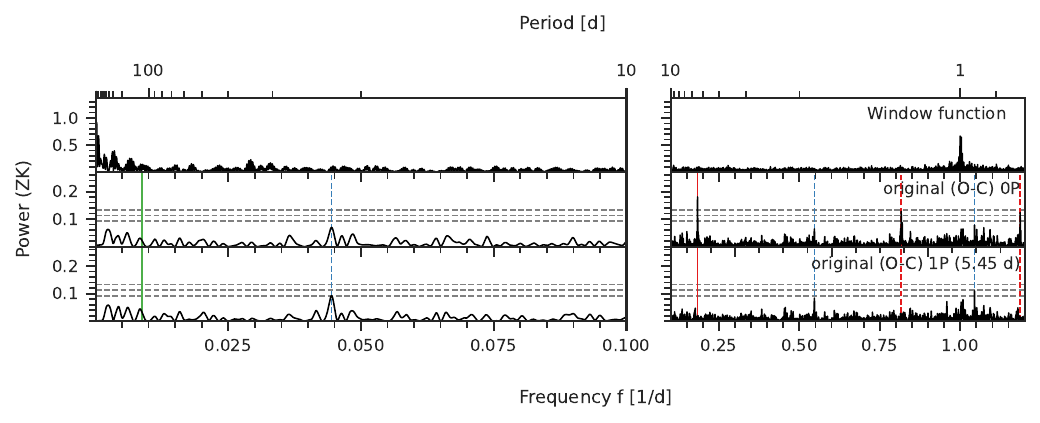}
    \caption{Same as Fig.~\ref{fig:gls_rv_J01048-181} but for \StarJOne{}.
    We indicate periods of the planet (5.45\,d), its aliases (\SI{1.22}{\day} and \SI{0.84}{\day}), and stellar rotation (\SI{114}{\day}) determined by \cite{Irwin2011}.
    Besides, the additional signals at periods of \SI{10.95}{\day}, \SI{1.83}{\day}, and \SI{22.49}{\day} are marked by blue dashed lines.}
    \label{fig:gls_rv_J19242+755}
\end{figure*}

Our intensive RV monitoring campaign of \StarJOne{} (GJ~1238, J19242+755) over a time span of two years revealed a small temperate planet, which orbits its host star with a period of $P=\PeriodJOne{}$ and a minimum mass of $M_{\rm pl}\sin i =\MassJOne{}$. The complete set of its model parameters is given in \autoref{tab:planetparams}, the phase-folded plot of the RVs to the single Keplerian model is illustrated in \autoref{fig:phasefolded_J19242+755}, and the RVs over time are plotted in \autoref{fig:rvs_over_time_J19242+755}.
As depicted in \autoref{fig:gls_rv_J19242+755}, its signal is highly significant in the GLS periodogram of the system, with FAP $< \num{2.8E-06}$.
Also prominent are further signals that we attributed to its daily aliases at periods of \SI{1.22}{\day} and \SI{0.84}{\day} with FAP $< \SI{0.17}{\percent}$ and FAP $< \SI{0.28}{\percent}$, respectively. These peaks are no longer notable in the residuals of the model after subtracting the planet signal. However, the residuals suggest three additional signals at periods of \SI{0.95}{\day}, \SI{22.49}{\day}, and \SI{1.83}{\day}. The most prominent of them, at \SI{0.95}{\day}, has a FAP below \SI{1.4}{\percent} and is related to the \SI{22.49}{\day} signal by daily sampling, while the \SI{1.83}{\day} signal with the lowest significance of FAP $< \SI{23}{\percent}$ appears to be independent of the others. 

We further investigated the consistency of these additional additional signals using stacked-Bayesian GLS (s-BGLS) periodograms \citep{sBGLS_lit}. They are depicted in \autoref{fig:sBGLS_J19242+755} and all of them show significant variability over time. 
Due to this incoherence and not being sufficiently significant in the GLS, we did not consider them in our model, but since they could turn out to be promising candidates, the system should be further monitored. More data will be helpful to reveal their true nature. 

For \StarJOne{}, we also explored the possibility that the additional signals could be spurious and induced by activity. As the analysis of possible period clustering (see \autoref{fig:activity_clusters_J19242+755}) and the GLSs of CARMENES activity indicators with signals of less than \SI{10}{\percent} FAP (see \autoref{fig:activity_GLS_J19242+755}) show, apparently none of the discussed periodic signals are related to the \ProtJOne{} rotation period of \StarJOne{} \citep{Irwin2011}, nor other forms of activity that would imprint periodic signals onto our RVs.

\subsection{Discovery of G~192--15\,b and G~192--15\,c}

\begin{figure*}[ht]
    \centering
    \includegraphics[width=0.93\textwidth]{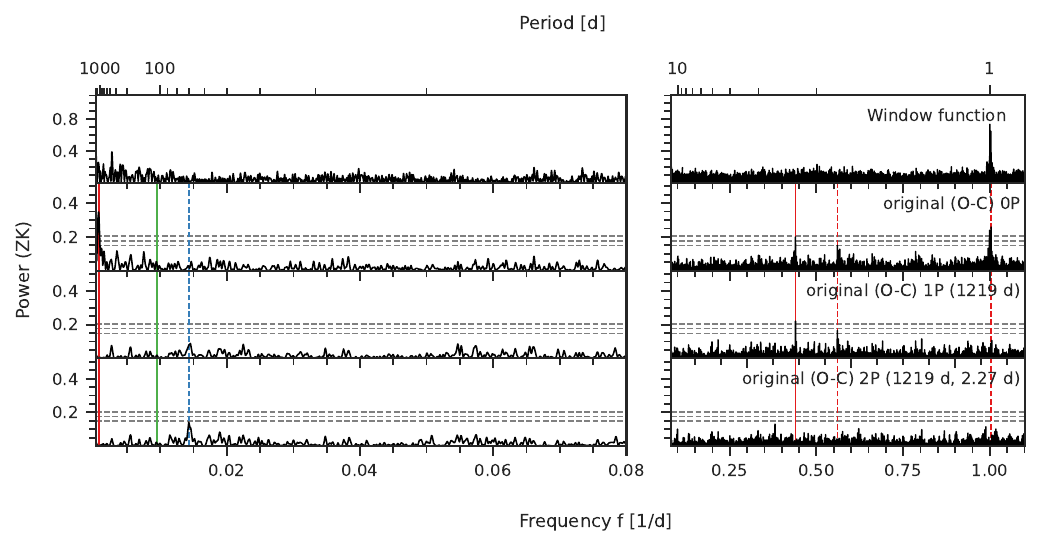} 
    \caption{Same as Fig.~\ref{fig:gls_rv_J01048-181} but for \StarJTwo{}.
    Extra panels at the bottom are for the GLS after subtracting the two-planet model.
    We indicate the periods of the two planets at \SI{1219}{\day} and \SI{2.27}{\day} (red solid lines), their aliases (red dashed lines), and stellar rotation (\SI{105}{\day}) determined by \citet{DiezAlonso2019} (green solid line).
    Besides, the additional signal at a period of \SI{69.9}{\day} is marked by the blue dashed line.}
    \label{fig:gls_rv_J06024+498}
\end{figure*}

The CARMENES RV measurements of G~192--15 (GJ~3380, J06024+498) collected over more than seven years reveal two substellar companions in the system. The GLS periodogram (see \autoref{fig:gls_rv_J06024+498}) shows a highly significant long periodic peak with an FAP $< \num{9.0E-10}$ at around \SI{1289}{\day} together with its daily alias at \SI{0.997}{\day} and FAP $< \num{7.2E-06}$. 
After modeling this signal with an eccentric orbit, a second significant signal in the residuals becomes apparent at a period of \SI{2.27}{\day} with an FAP $< \num{2.2E-04}$. The one-day alias of this secondary peak at \SI{1.78}{\day} is also evident, but of less significance. We tested different models with varying number of planets, as well as allowing for non-zero eccentricities, and concluded that there are two planets orbiting the stellar host G~192--15. The evidences for the various models are tabulated in \autoref{tab:modelcomp_G192-015}.

\begin{table}
\caption{Model comparison for G~192--15.}
\label{tab:modelcomp_G192-015}
\centering
\begin{tabular}{l S[table-format=-3.1] S[table-format=-2.1]}
\hline\hline
    \noalign{\smallskip}
               Model &  {$\ln\mathcal{Z}$} &  {$\Delta\ln\mathcal{Z}$} \\
    \noalign{\smallskip}
\hline
    \noalign{\smallskip}
0P &            -426.0 &                -37.8 \\
1P$_\text{(1219 d-circ)}$ &            -404.6 &                 -16.4  \\
1P$_\text{(1219 d-ecc)}$ &            -398.8 &                 -10.6  \\
\noalign{\medskip}
2P$_\text{(1219 d-circ, 2.27 d-circ)}$ &            -396.5 &                -8.3 \\
2P$_\text{(1219 d-ecc, 2.27 d-circ)}$ &            -388.2 &                0 \\
2P$_\text{(1219 d-ecc, 2.27 d-ecc)}$ &            -391.3 &                -3.1 \\
\noalign{\medskip}
3P$_\text{(1219 d-ecc, 2.27 d-circ, 70 d-circ)}$ &            -386.6 &                1.6 \\
3P$_\text{(1219 d-ecc, 2.27 d-circ, 70 d-ecc)}$ &            -389.6 &                -1.4 \\
    \noalign{\smallskip}
\hline
\end{tabular}
\end{table}

\begin{figure*}[!tp]
    \centering
    \includegraphics[width=0.44\textwidth]{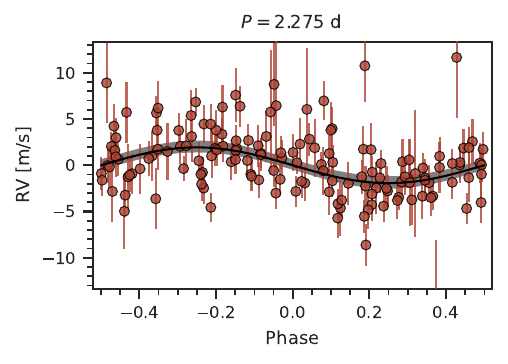} 
    \includegraphics[width=0.44\textwidth]{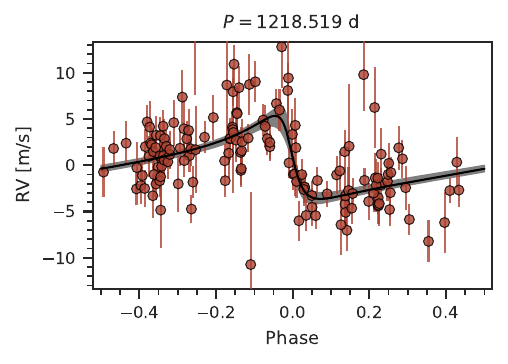}
    \includegraphics[width=0.88\textwidth]{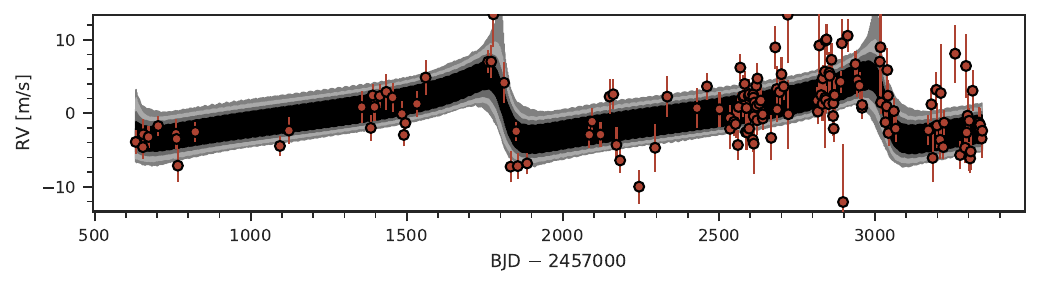}
    \caption{Upper panel: Same as Fig.~\ref{fig:phasefolded_J01048-181} but for the simultaneous best fit model ($=\text{2P}_\text{(1218\,d-ecc, 2.27\,d-circ)}$) of the two planetary signals around \StarJTwo. Lower panel: RVs over time for the same model. The black lines show the model based on the parameters listed in \autoref{tab:planetparams}, and the gray shaded areas denote the \SI{68}{\percent}, \SI{95}{\percent,} and \SI{99}{\percent} confidence intervals, respectively. The instrument offset of CARMENES was subtracted from the measurements and the model.}
    \label{fig:phased_RVs_J06024}
\end{figure*}

The long period planet exhibits a high minimum mass of $M_{\rm pl}\sin i =\MassJTwoc{}$ and revolves around its host star on an eccentric orbit with $e \approx 0.68$. For the short-period planet we derived a minimum mass of $M_{\rm pl}\sin i =\MassJTwob{}$ and our model comparison favors a circular Keplerian over an eccentric one. The complete set of model parameters are listed in \autoref{tab:planetparams}, and the phase-folded RVs together with the entire RV time series can be found in \autoref{fig:phased_RVs_J06024}.

After subtracting our two-planet solution from the RVs, we found another signal in the residuals at a period of \SI{69.94}{\day}. While it does not reach a high level of significance (FAP around $\SI{15}{\percent}$), a three-planet solution appears at least fairly reasonable with $\Delta\ln\mathcal{Z} = 1.6$ against our preferred model. For further investigation, we examined all the signals' stability with increasing number of observation. As depicted in the s-BGLSs in \autoref{fig:sBGLS_J06024}, the two planetary signals appear coherent and increase in significance over time, whereas the third candidate signal shows some variability. In addition to that, we used \texttt{Exostriker} \citep{Trifonov_exo_2019} to run an orbital evolution for such a three-planet configuration and found that, while the system is stable for the two planets alone, they would undergo significant changes in their orbital eccentricities in the presence of the third companion. For those reasons, we did not consider the third signal any further, but again additional monitoring of the system would be helpful for its understanding. 

We also analyzed the activity indices with the most significant signals, as we did for the previous systems. The corresponding GLS periodograms (\autoref{fig:activity_GLS_J06024}), however, do not reveal any significant power at the periods of the discovered companions, nor at the rotation period of the star at \SI{105}{\day} \citep{DiezAlonso2019}.

\section{Planet occurrences}
\label{subsubsec: detcompleteness}

\begin{table*}
\caption{Known planets around stars in our sample.}
\label{table:planets}   
\centering 
\begin{tabular}{
  ll
  >{\collectcell\num}r<{\endcollectcell}
  @{${}\pm{}$}
  >{\collectcell\num}l<{\endcollectcell}
  >{\collectcell\num}r<{\endcollectcell}
  @{${}\pm{}$}
  >{\collectcell\num}l<{\endcollectcell}
  >{\collectcell\num}r<{\endcollectcell}
  @{${}\pm{}$}
  >{\collectcell\num}l<{\endcollectcell}
  l
}

\hline\hline                
\noalign{\smallskip}
Karmn & Planet ID & \multicolumn{2}{c}{$P_\text{pl}$} & \multicolumn{2}{c}{$M_\text{pl} \sin i$} & \multicolumn{2}{c}{$a_\text{pl}$} & References \\
~  & ~ &  \multicolumn{2}{c}{[d]}  & \multicolumn{2}{c}{[M$_\oplus$]} & \multicolumn{2}{c}{[au]} & ~ \\
\noalign{\smallskip}
\hline                                   
\noalign{\smallskip}
J00067--075\tablefootmark{{a}}	& G~158--27\,b & 10.347 & 0.027  & 0.99 & 0.12 & 0.0457&0.0013 & SMa23 \\
            & G~158--27\,c & 21.20 & 0.13 &  1.25 & 0.16  & 0.0738 & 0.0021 & SMa23 \\
J01048--181	& G~268--110\,b &  1.432630 & 76e-6  &  1.52 & 0.25  &  0.01283 & 0.00029    & This work \\
J01125--169	& YZ~Cet\,b &  2.020870& 90e-6 	  &  0.686 & 0.088 	&  0.01634 & 0.00041  & Ast20, Sto20 \\
            & YZ~Cet\,c &  3.05989 & 0.00010  &  1.12 & 0.11    &  0.02156 & 0.00054  & Ast20, Sto20 \\
            & YZ~Cet\,d &  4.65626 & 0.00029  &  1.07 & 0.12 	&  0.02851 & 0.00071  & Ast20, Sto20 \\
J02530+168	& Teegarden's Star\,b &  4.90634 & 0.00041     &	 1.16 & 0.12  &  0.0259 & 0.0009  & Zec19, Dre24 \\
            & Teegarden's Star\,c &  11.4160 & 0.0030   &	 1.05 & 0.14 	&  0.0455 & 0.0016  & Zec19, Dre24 \\
            & Teegarden's Star\,d &  26.130 & 0.040   &	 0.82 & 0.17 	&  0.0791 & 0.0027  & Zec19, Dre24 \\
J03133+047	& CD~Cet\,b &  2.29070 &	0.00012 	  &  3.95 & 0.43 	&  0.0185 & 0.0013  & Bau20 \\
J06024+498	& G~192--15\,b &  2.27476& 0.00028 	  &  1.03 & 0.18 	&  0.01723 & 0.00039  & This work \\
            & G~192--15\,c &  1219 & 13  &  14.3 & 1.6    &  1.137 & 0.027  & This work \\
J08413+594	& G~234--45\,b &  203.59& 0.14 	  &  146.7 & 7.0 	&  0.3380 & 0.0084  & Mor19, Rib23 \\
            & G~234--45\,c &  2350 & 100  &  143.0 & 7.0    &   1.722 & 0.049  & Rib23 \\
J19242+755 	& G~261--6\,b &  5.4536 & 0.0032  &	 1.37 & 0.23   &  0.02971 & 0.00093  & This work \\
J20260+585 	& Wolf~1069\,b &  15.564 & 0.015  &	 1.22 & 0.20  &  0.0672 & 0.0014  & Kos23 \\
\noalign{\smallskip}
\hline                                         
\end{tabular}
\tablefoot{SMa23: \cite{Mascareno23}; Ast20: \cite{Astudillo2017A&A...605L..11A}; Sto20: \cite{Stock2020a}; Zec19: \cite{Zechmeister2019}; Dre24: \cite{Dreizler2024}; Bau20: \cite{Bauer2020}; Mor19: \cite{Morales2019}; Rib23: \cite{Ribas2023}; Kos23: \cite{Kossakowski2023}. \tablefoottext{a}{Planets around this host star can not be identified with CARMENES data alone.}}
\end{table*}

Planet occurrence rates are estimated from the number of detected planets within particular mass or period bins around a sample of investigated stars. The known planets around the stars from the investigated stellar sample, together with the ones presented in this work, are listed in \autoref{table:planets}. Wherever applicable, the listed planetary masses are rescaled by stellar masses updated since the original publications, as provided in \autoref{table:stellar_properties}. 

Since the planet candidates proposed by \citet{Mascareno23} around G~158--27 (GJ~1002, J00067--075) can not be identified from the CARMENES RVs alone, they were not taken into account in the further statistical analysis regarding planet occurrences. Beside that, as in this work we studied only companions with orbital periods of up to 100\,d and planetary masses below 10\,M$_\oplus$, any known planets above these limits were also omitted, which left us with 11 planets in seven systems (i.e. all in \autoref{table:stellar_properties} except for J00067--075 b and c, J06024+498 c, and J08413+594 b and~c).

The numbers of planet discoveries are generally affected by the capabilities of the instruments used for the observations of the targets, as well as the methods applied for data reduction and eventually signal and planet detection. Naturally, unknown planets can be missed in RV surveys, which if uncorrected, can lead to a bias and an underestimation of the final occurrence rates. To overcome this issue, we followed the procedures described by \cite{Sabotta2021} and estimated the planet detection completeness within our stellar sample by means of an injection-and-retrieval analysis similar to those previously used~\citep{Cumming_1999, Zechmeister2009, Meunier_2012, Bonfils2013, Wolthoff2022}.

\subsection{Preprocessing RV data}\label{subsec: RVsignals}

As it is preferable to perform the injection-and-retrieval analysis for the estimation of the detection completeness (see \autoref{subsec:Det_completeness}) on clean RV data, we applied a prewhitening procedure on the time series contained in the analyzed sample. To do so, as in \cite{Sabotta2021}, we first performed 3$\sigma$ outlier clipping and computed GLS periodograms for all targets. We then repeatedly fitted Keplerian orbits to signals with a FAP of 1\,\% or lower, until no such signals were left.
In this manner we retrieved the signals of the known planets, but also of others that we attributed either to activity, stellar rotation, or telluric contamination. 
Activity signals were identified by using the time series of the activity indicators evaluated by {\tt serval}. 
On the other hand, if the signals' periods coincide with known stellar rotation periods or their first high harmonics, we attributed them to stellar rotation. Finally, systematics due to telluric contamination were identified by comparing the RV data with those determined from the same spectra after the correction for telluric absorption following \cite{Nagel_2023}. However, for some of the found signals, their origin cannot be settled with the current data available to us. All signals found and identified following the described approach are listed in Table~\ref{table:signals}.

\begin{table}
\caption{Removed RV signals before injection-and-retrieval.}
\label{table:signals}   
\centering 
\begin{tabular}{l c l}   
\hline\hline                
    \noalign{\smallskip}
Karmn & $P_{\rm pl}$ [d] & Remark \\
    \noalign{\smallskip}
\hline                                   
    \noalign{\smallskip}
J00184+440   & 307.7   &  Probably activity \\ 
             & 93.99   &  Probably tellurics \\
J01048--181  & 1.43   &  Planet G~268--110\,b\\ 
J01125--169  & 2.01   &  Planet YZ~Cet\,b \\ 
             & 3.06   &  Planet YZ~Cet\,c \\ 
             & 4.652   &  Planet YZ~Cet\,d \\ 
             & 81.0   &  Activity \\
J02530+168   & 4.9   &  Planet Teegarden's Star\,b \\ 
             & 11.4   &  Planet Teegarden's Star\,c \\ 
             & 26.1   &  Planet Teegarden's Star\,d \\ 
             & 96.16   &  Rotation \\
             & 174.4   &  Telluric contamination \\
             & 2949.8   &  Probably activity \\
J03133+047   & 2.29   &  Planet CD~Cet\,b \\ 
             & 67.9   &  Probably rotation \\
J06024+498   & 2.27   &  Planet G~192--15\,b \\ 
             & 1213.7   &  Planet G~192--15\,c \\ 
J08413+594   & 203.14   &  Planet G~234--45\,b \\ 
             & 2354.3   &  Planet G~234--45\,c \\ 
J18027+375   & 97.41   &  Unsolved \\
J19242+755   &   5.45  &  Planet G~261--6\,b \\ 
J20260+585   & 396.39   &  Probably tellurics \\ 
             & 15.54   &  Planet Wolf 1069\,b \\ 
             & 146.74   &   Unsolved, probably tellurics? \\ 
J23419+441   & 178.87   &  Telluric contamination \\ 
             & 93.99   &  Probably tellurics \\
\noalign{\smallskip}
\hline 
\end{tabular}
\end{table}

\subsection{Detection completeness}\label{subsec:Det_completeness}

\begin{figure}
    \centering
    \includegraphics[width=\linewidth]{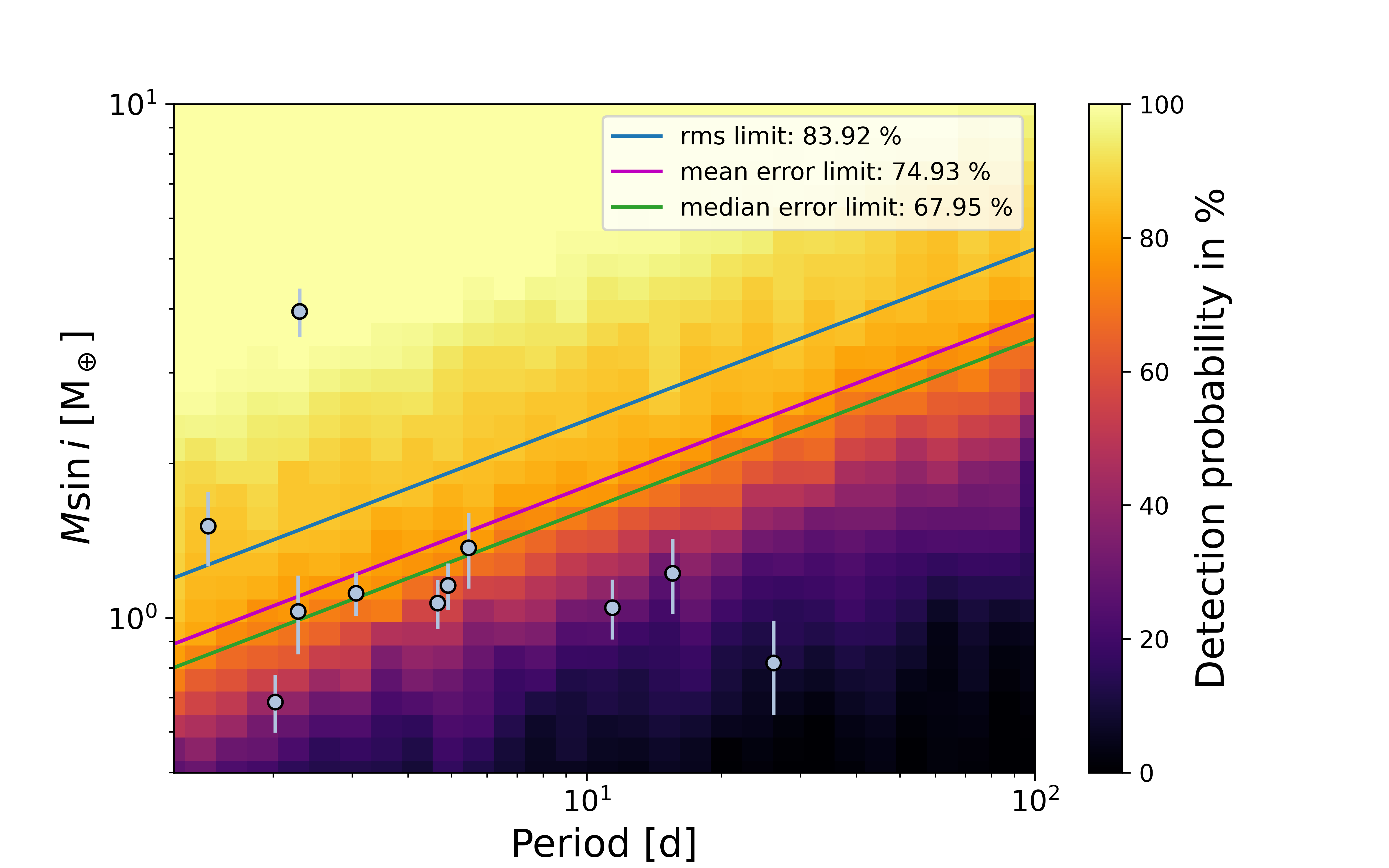}
    \caption{CARMENES detection sensitivity map, averaged over the individual maps of the 15 stellar targets of this study. 
    The light blue markers indicate the 11 planets included for this occurrence rate analysis, and the color map illustrates the detection probabilities of the respective period-mass grid points. The solid lines represent the masses associated to the RV semi-amplitude values equal to the RVs' root mean square averaged over the time series (blue), the mean RV uncertainties (magenta), and the median of the RV uncertainties.}
    \label{fig:planet_occurrence}
\end{figure}

In order to estimate the detection sensitivity within our data sets, we injected single artificial planets on circular orbits into our RV data and tested whether we were able to retrieve them. In this step, we neglected eccentricity, since it introduces an additional complexity to the problem, but, as implied by \cite{Cumming2004} and verified by \cite{Sabotta2021}, does not significantly change the outcome. We counted individual tests as successful recoveries if the signal in question appears as the highest peak in the resulting GLS periodogram with a FAP below \SI{1}{\percent}. We did this 200 times for all combinations of orbital periods $P$ and planet minimum masses $M_\mathrm{pl} \sin i$ spanned by a log-uniform grid of 30 points in the period range 1--100\,d and 30 grid points in the planet minimum mass range 0.5--10\,M$_\oplus$, i.e., 180\,000 different simulations in total.
For all these planetary signals, the phase $\phi$ was chosen at random, and the semi-amplitude was determined by the common approximation

\begin{equation} \label{eq:semi_amplitude}
K=\mathrm{\SI{28.435}{\meter\per\second}} \left(\frac{P}{1\,\mathrm{yr}} \right)^{-1/3}
\left(\frac{M_\mathrm{pl}\sin{i}}{M_\mathrm{Jup}} \right)\left(\frac{M_\star}{M_\odot} \right)^{-2/3} \,.
\end{equation}
Consequently, the RV signals of the artificial planets
\begin{equation} \label{eq:RV_signal}
RV(t)=K \sin{\left(\frac{2\pi t}{P} + \phi \right)}
\end{equation}

\noindent were evaluated at the given time stamps of the observations and added up onto the RV time series after they had been prewhitened, as described in \autoref{subsec: RVsignals}.
The individual detection maps on the chosen grid of parameters for each tested star were next determined by the fraction of successfully recovered planetary signals. After averaging those detection maps over all of the 15 stellar targets, we obtained the final detection sensitivity map of the survey's subsample, which is plotted in \autoref{fig:planet_occurrence}.

\subsection{Planet occurrence rates}\label{subsubsec: occrates}

We determined the final occurrence rates within the period range 1--100\,d and minimum mass range 0.5--10\,M$_\oplus$ by running a Monte Carlo simulation on a grid of planet frequencies $\overline{n}_\text{pl}$, in terms of number of planets per star, to test how consistent it is with the actual number of detected planets $N_\mathrm{pl,det}$. The simulation was run 1000 times iteratively for each of the test frequencies and consisted of the following steps:

\begin{enumerate}[(a)]
\item Given the test frequency $\overline{n}_\text{pl}$, a test planet sample is created, for which the number of test planets $N_\mathrm{pl,in}$ is determined by the Poisson distribution $\lambda=\overline{n}_\text{pl} N_\star$, where $N_\star$ is the number of stars within our underlying stellar sample.

\item A period and minimum mass from the mass-period grid of the detection sensitivity map are assigned to each of the test planets. We applied a log-uniform distribution in the period, whereas for the mass we used a power-law distribution of the form $N_\mathrm{pl}=a(M_\mathrm{pl}\sin{i})^\alpha$, with $a=319.25$ and $\alpha=-1.06$ \citep{Ribas2023}. To test the robustness of the occurrence rate estimations to the assumption of the underlying planet mass distribution, we alternatively applied a log-uniform distribution in mass as well.

\item The number of test planet retrievals $N_\mathrm{pl,out}$ is determined by the count of detections within the sample of the test planets. Here, each of the test planets was accepted as a successful detection based on a random draw with a binary chance of success according to its corresponding detection probability.
\end{enumerate}

For each of the test frequencies $\overline{n}_\text{pl}$ we counted how often, out of the 1000 simulation runs, the number of retrieved test planets equals the actual number of known detected planets $N_\mathrm{pl,out} = N_\mathrm{pl,det}$. Based on these counts, the resulting probability density over the given grid of tested frequencies was finally normalized, and from the locations of its median, as well as its 16th and 84th percentiles, we derived the most probable frequency (occurrence rate) and its corresponding uncertainties, as illustrated by \autoref{fig:occ_rates_powerlaw}.

\section{Results and discussion}\label{sec: discussion}

\subsection{Late M dwarfs host Earth-like planets and very few super-Earths}

\begin{table}
\caption{Occurrence rates $\overline{n}_\text{pl}$ in terms of planets per star from the 15 target stars based on sensitivity maps from the injection-and-retrieval analysis and based on the sensitivity map derived from likelihood maps.}
\label{tab:occurrence_all_merged}   
\centering
\begin{tabular}{l c c c c}   
\hline\hline     
\noalign{\smallskip}
& \multicolumn{2}{c}{Inj.-and-retrieval} & \multicolumn{2}{c}{Log-likelihood} \\
\noalign{\smallskip}
\multicolumn{1}{c}{$M_\mathrm{pl} \sin i$} &$P$ [d]&$P$ [d]&$P$ [d]&$P$ [d] \\
&1--10&10--100 &1--10&10--100 \\                                
\noalign{\smallskip}
\hline   
\noalign{\smallskip}
\multicolumn{5}{c}{power-law distribution} \\
\noalign{\smallskip}
\hline                                   
\noalign{\smallskip}
0.5\,M$_\oplus$--3\,M$_\oplus$ & 0.88$^{+0.36}_{-0.28}$&0.92$^{+0.56}_{-0.39}$& 0.77$^{+0.31}_{-0.24}$&1.00$^{+0.61}_{-0.43}$\\ 
\noalign{\smallskip} 
3\,M$_\oplus$--10\,M$_\oplus$ & 0.11$^{+0.11}_{-0.06}$&0.06$^{+0.09}_{-0.04}$& 0.12$^{+0.11}_{-0.07}$&0.06$^{+0.08}_{-0.04}$  \\ 
\noalign{\smallskip}
\hline  
\noalign{\smallskip}
\multicolumn{5}{c}{log-uniform distribution} \\
\noalign{\smallskip}
\hline                                   
\noalign{\smallskip}
0.5\,M$_\oplus$--3\,M$_\oplus$ & 0.72$^{+0.29}_{-0.23}$&0.63$^{+0.38}_{-0.27}$ & 0.65$^{+0.26}_{-0.21}$&0.67$^{+0.40}_{-0.29}$\\ 
\noalign{\smallskip} 
3\,M$_\oplus$--10\,M$_\oplus$ & 0.11$^{+0.11}_{-0.07}$&0.06$^{+0.08}_{-0.04}$ & 0.12$^{+0.11}_{-0.07}$&0.05$^{+0.08}_{-0.04}$ \\ 
\noalign{\smallskip}
\hline                                            
\end{tabular}
\tablefoot{Upper: Under the assumption of a power-law distribution in $M_\text{pl} \sin i$ from \cite{Ribas2023}. Lower: Under the assumption of a log-uniform distribution in $M_\text{pl} \sin i$. The error bars show the 16\,\% and 86\,\% levels of the resulting distribution.}
\end{table}

\begin{figure*}[ht]
    \centering
    \includegraphics[width=\linewidth]{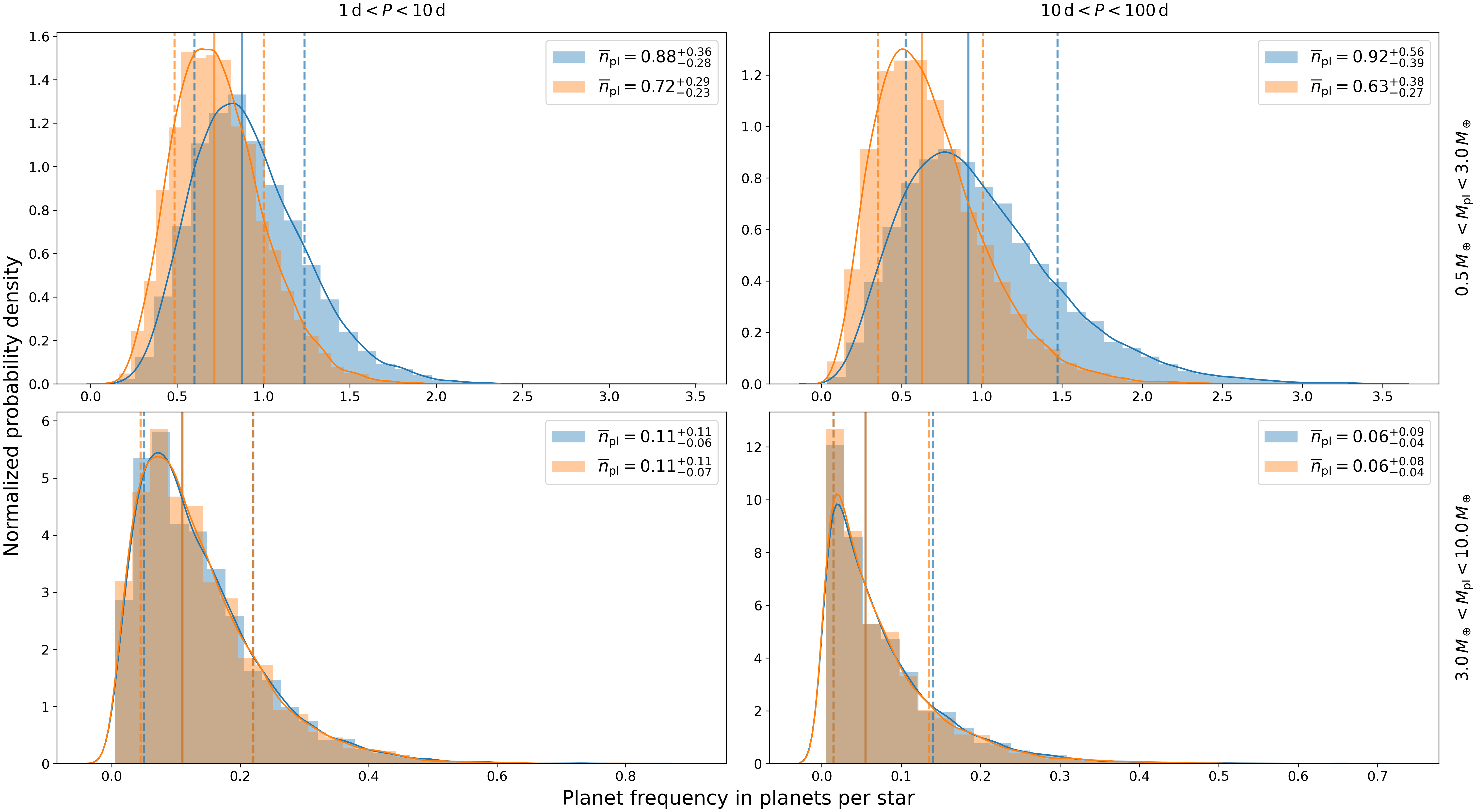}
    \caption{Occurrence rate distributions for different combinations of period and planetary mass ranges. The medians as well as the 16th and 84th percentiles are indicated by vertical solid and dashed lines, respectively. The distributions in blue are based on an underlying power-law distribution for the planet masses, while the orange ones are estimated using a log-uniform distribution.}
    \label{fig:occ_rates_powerlaw}
\end{figure*}

\begin{figure}
    \centering
    \includegraphics[width=\linewidth]{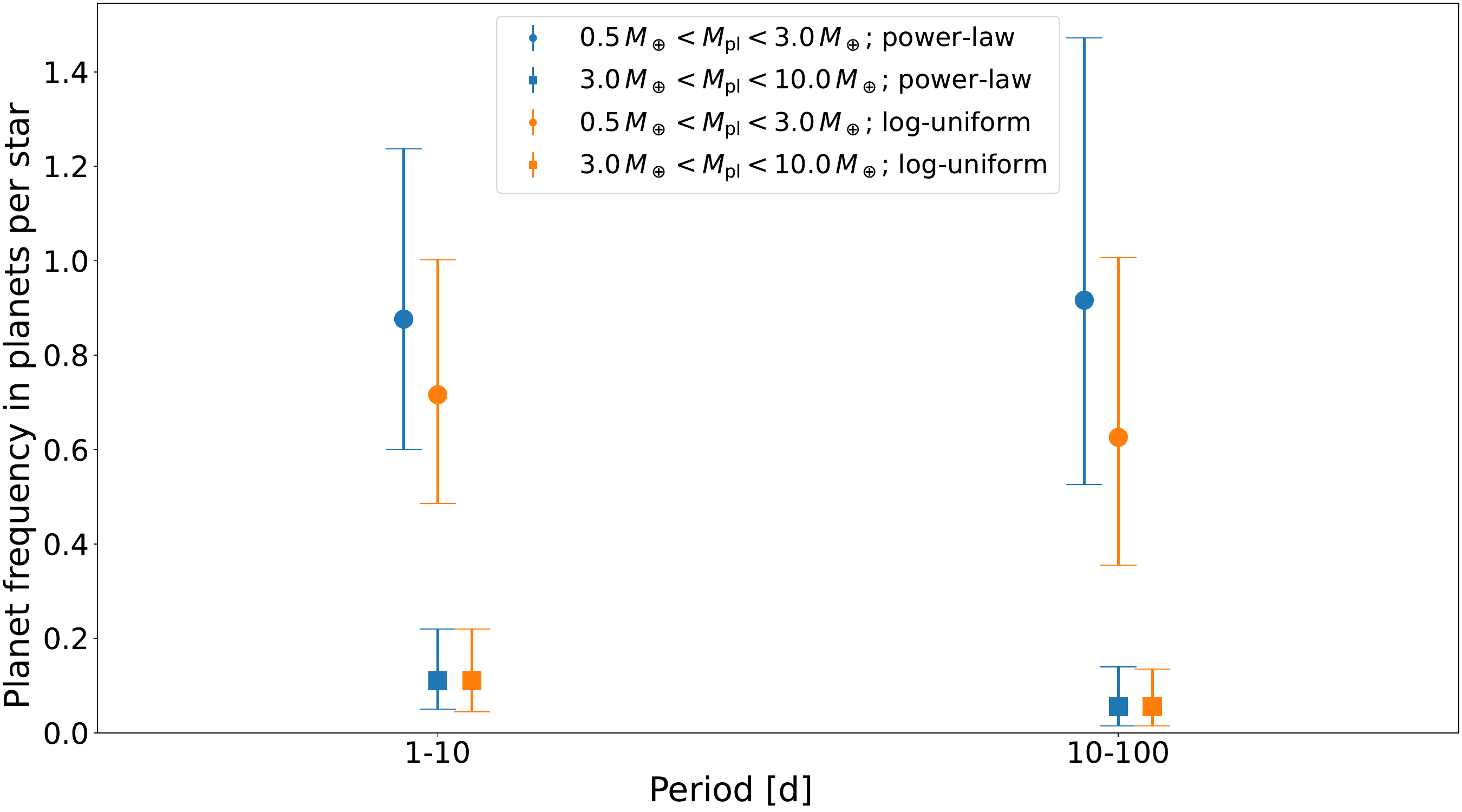}
    \caption{Occurrence rates based on the power-law distribution in planetary masses with respect to orbital periods. The different bins in planetary masses are color and symbol coded.}
    \label{fig:occ_rates_powerlaw2}
\end{figure}

We determined the occurrence rates for four distinct, not overlapping bins in the mass-period plane with orbital periods $P$ between 1\,d and 100\,d and minimum planetary masses $M_\mathrm{pl} \sin i$ between 0.5\,M$_\oplus$ and 10\,M$_\oplus$. They are listed in \autoref{tab:occurrence_all_merged}, and the underlying distributions are plotted in \autoref{fig:occ_rates_powerlaw}. Our results indicate a significant dependency on planetary masses. Whereas the occurrence rates for low planetary masses below 3\,M$_\oplus$, namely $\overline{n}_\text{pl} = 0.88^{+0.36}_{-0.28}$ for short orbital periods below $P=10$\,d and $\overline{n}_\text{pl} = 0.92^{+0.56}_{-0.39}$ for periods above $P=10$\,d, agree well with one planet per star, they decrease substantially for planetary masses between 3\,M$_\oplus$ and 10\,M$_\oplus$ to $\overline{n}_\text{pl} = 0.11^{+0.11}_{-0.06}$ and $\overline{n}_\text{pl} = 0.06^{+0.09}_{-0.04}$, respectively. This tendency is illustrated in \autoref{fig:occ_rates_powerlaw2}. 
For planets below 3\,M$_\oplus$ the occurrence rates that we obtained appear to be constant with respect to orbital periods. Therefore, we cannot confirm the implications from \cite{Sabotta2021}, who found increased occurrence rates for orbital periods below $10$\,d for stellar hosts of masses below $0.34$\,M$_\odot$, which is twice as high as the mass threshold for the stellar sample that we investigated. While our occurrence rates for the planetary mass range between 3\,M$_\oplus$ and 10\,M$_\oplus$ may indicate such a trend, with only one known planet within this mass range, the statistics in this regime are not particularly reliable. Consequently, the corresponding uncertainties are high, and the values obtained are mutually consistent within their errors.

Although the occurrence rates are consistently lower when using a log-uniform distribution for the planetary masses, they agree within their uncertainties to those estimated from the power-law distribution, and the relations with periods and masses prevail. However, the rates for planets with masses below 3\,M$_\oplus$ using the power-law distribution appear to be significantly increased with respect to those based on the log-uniform distribution when wider orbits are considered. This is due to an increased number of low-mass planets on wide orbits in the simulated planet samples. As those low-mass planets have small RV semi-amplitudes, they naturally fall into regions of low detection probabilities. This in turn leads to higher, possibly overestimated occurrence rates. A more realistic and applicable distribution of planetary masses with a dependence on the orbital period would be helpful to compensate for that possible bias. 

All in all, our determined occurrence rates are increased for smaller planet masses for any of the investigated planetary orbit regimes. In fact, the most massive planet in our sample, with $M_\mathrm{pl} \sin i = 3.95 \pm 0.43$\,M$_\oplus$, revolves around the most massive stellar host within our sample, namely CD~Cet (J03133+047), with a mass of 0.161\,M$_\odot$.
In addition, the comparison to the occurrence rates from \cite{Ribas2023}, which were derived from the complete CARMENES stellar sample at the time, indicates a significant dependence of planetary rates on spectral type, even within the M-dwarf regime alone. For planetary masses between 1\,M$_\oplus$ and 10\,M$_\oplus$, the authors reported significantly lower occurrence rates of $\overline{n}_\text{pl} = 0.39^{+0.10}_{-0.07}$ for short orbits below $10$\,d than for wider orbits up to $100$\,d, with $\overline{n}_\text{pl} = 0.67^{+0.18}_{-0.15}$. In contrast to that, the results from the present work, for which only the least massive stars were considered, indicate a significantly increased number of planets on short orbits up to $10$\,d and for planetary masses from 0.5\,M$_\oplus$ to 10\,M$_\oplus$. For that same short-period regime we consequently arrive at an occurrence rate of $\overline{n}_\text{pl} = 0.99$ planets per star, which is at least twice as high as that reported by \cite{Ribas2023}. This number can be derived from the combination of the rates for low-mass planets below 3\,M$_\oplus$ $\overline{n}_\text{pl} = 0.88$, and $\overline{n}_\text{pl} = 0.11$ for masses between 3\,M$_\oplus$ and 10\,M$_\oplus$, as they represent independent bins of planet masses. This dependence on stellar masses is also evident when our rates are compared with those in other studies. For low-mass planets up to 10\,M$_\oplus$ and on orbits between $P=1$\,d and $P=100$\,d, \cite{Bonfils_2013_sample} found a rate of $\overline{n}_\text{pl} = 0.36^{+0.24}_{-0.10}$ for their HARPS M-dwarf sample with a median stellar mass of 0.33\,M$_\odot$, which is in line with \cite{Ribas2023}. An even lower rate of $\overline{n}_\text{pl} = 0.10^{+0.10}_{-0.03}$ was derived by \cite{Pinamonti_2022} for an even earlier stellar sample of M dwarfs with types between M0 and M3 and planets on orbits shorter than $P=10$\,d. Those trends again suggest that the number of small planets on short orbits is increased for late spectral types.

\begin{figure}
    \centering
    \includegraphics[width=\linewidth]{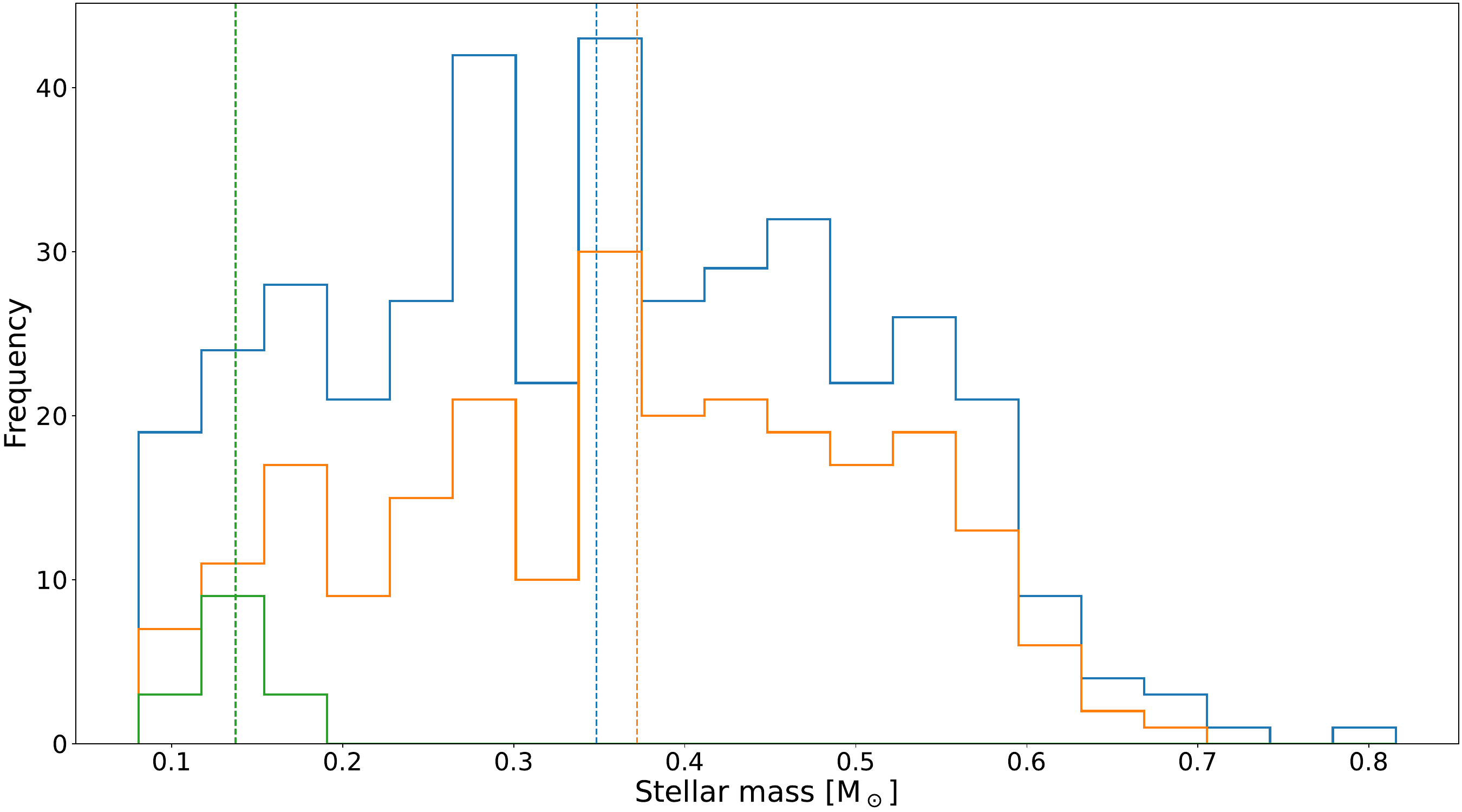}
    \caption{Distributions of stellar masses for our stellar sample (green), the subsample used by \cite{Ribas2023} for their occurrence rate analysis (orange), and the whole CARMENES GTO sample (blue). The median stellar masses, indicated by the vertical lines, are 0.137\,M$_\odot$, 0.372\,M$_\odot$, and 0.348\,M$_\odot$, respectively.}
    \label{fig:mass_samples_hist}
\end{figure}

To illustrate the difference of the analyzed samples, in \autoref{fig:mass_samples_hist} the distribution of stellar masses of our sample is plotted and put into context to the subsample from \cite{Ribas2023} and to the entire CARMENES sample. While the subsample from \cite{Ribas2023} represents the entire GTO sample quite accurately, the stars used for the current work are significantly less massive. We presume that with a sufficiently large pool of targets and observations, it will be possible to further resolve trends in planet demographics and their dependence on host star mass, clarifying the role of host star properties on the planet formation process.

Still, one should take note of the high uncertainties of the estimated rates, which are due to the small number statistics. The analyzed sample contains only 15 stars and 11 already detected planets that lie within our range of period of interest and that could be identified with CARMENES RVs alone. Also, although we tried to compensate by our methods for the incompleteness of the sample, it cannot be ruled out that the number of low-mass planets on wide orbits is still underestimated, as the detection sensitivity for Earth-like planets is still fairly low, even for orbits at around $P=20$\,d (see the sensitivity map in \autoref{fig:planet_occurrence}). 

For the sake of completeness, we note that there may be a small number of false-negative non-detections in RV surveys for planets on orbits with periods that are close to the stellar rotation periods of their hosts. Signals of such candidates tend to be falsely discarded, as the data often show concurrent stellar activity. From transit surveys the percentage of such cases can be estimated to be around 5\% (Lechuga et~al. in~prep.) and therefore we conclude that the planet occurrence rates will be underestimated only insignificantly.

\subsection{Detection probabilities from likelihood maps}

\begin{figure*}[!tp]
    \centering
    \includegraphics[width=0.49\textwidth]{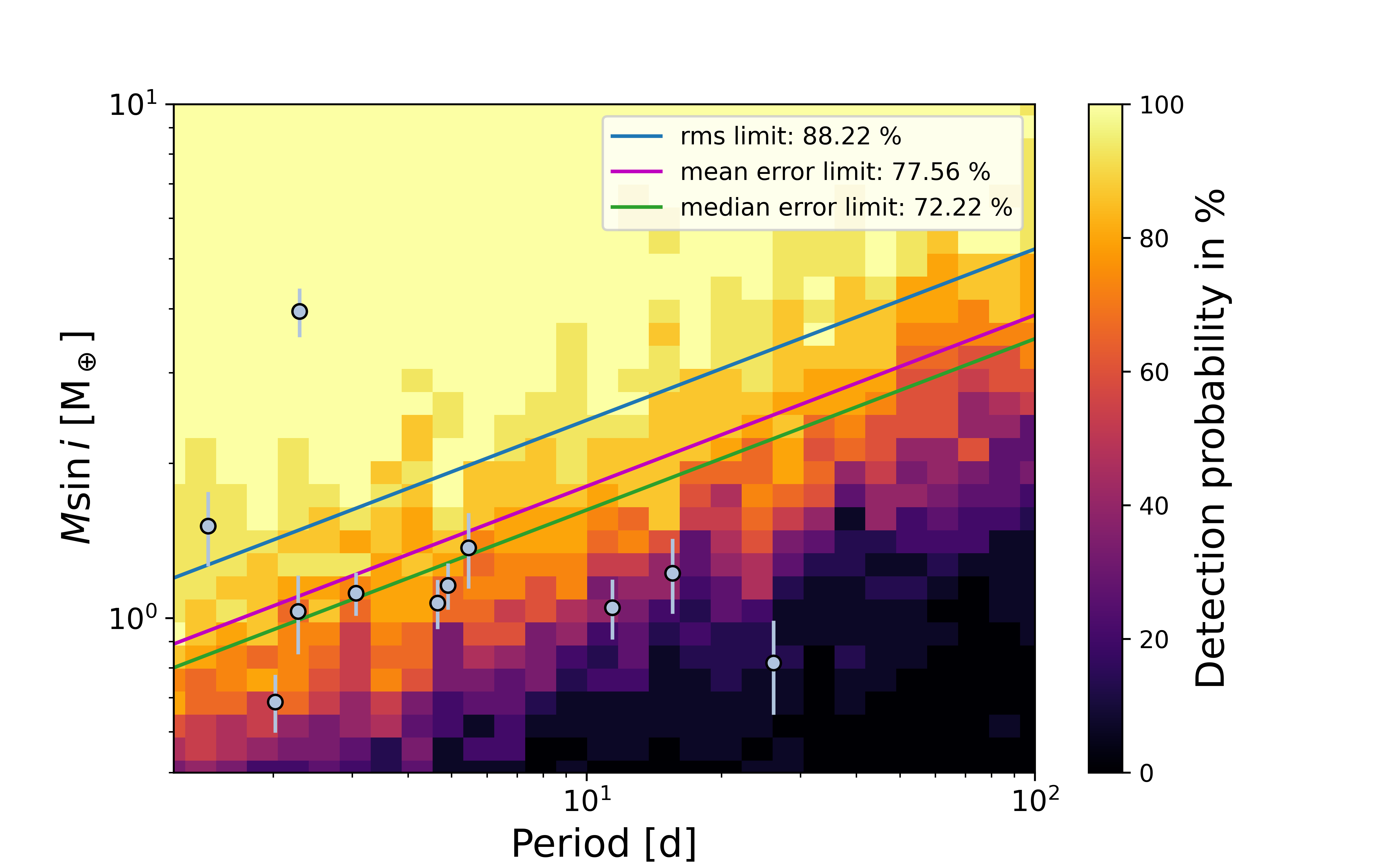} 
    \includegraphics[width=0.49\textwidth]{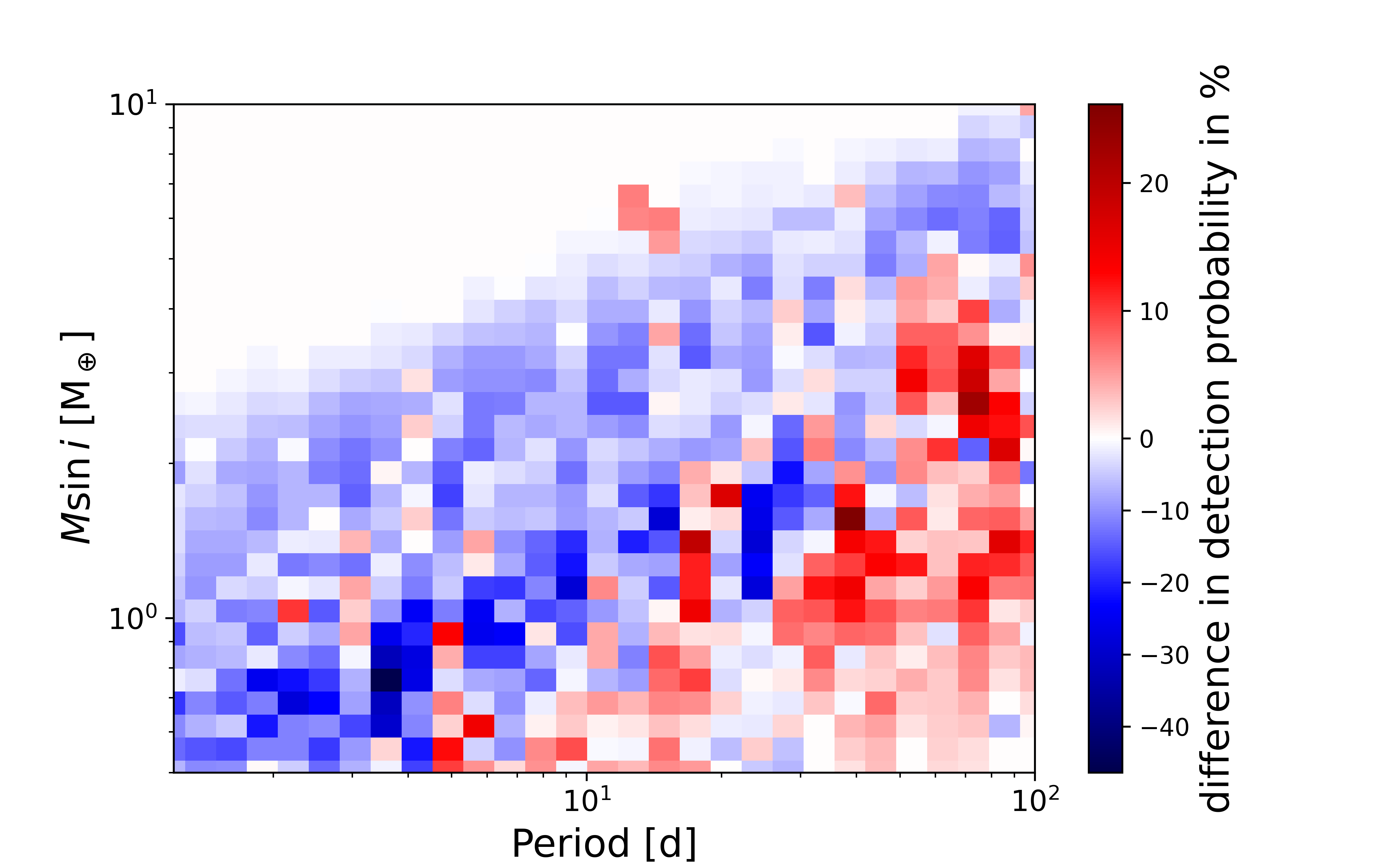}
    \caption{\textit{Left}: Same as \autoref{fig:planet_occurrence} but based on log-likelihood estimation and averaged over the individual binary maps of the 15 targets. \textit{Right}: Delta $\mathcal{P}_\text{inj.-retr.}-\mathcal{P}_\text{lik.}$ between the CARMENES detection sensitivity maps based on injection-and-retrieval vs. the one based on log-likelihood estimation.}
    \label{fig:sensmaps_loL_and_comp}
\end{figure*}

As the accurate estimation of the occurrence rates is heavily dependent on a proper characterization of the detection limits, we tested another method to derive the sensitivity map for the sample. For this we followed a variation of the approach introduced by \cite{Tuomi2014} and later applied also by \cite{Pinamonti_2022}. 
Whereas they used posterior sampling for the estimation of the detection probabilities, we relied on log-likelihood maps in order to identify the area on the plane spanned by planetary masses and orbital periods, where a planetary signal with the corresponding semi-amplitude would likely be detected. To achieve that, for each of our cleaned and prewhitened RV time series (see \autoref{subsec: RVsignals}), we estimated the likelihoods of the best fits of models with an additional postulated circular Keplerian signal. This was done on a two-dimensional grid for different orbital periods and planetary masses (RV semi-amplitudes), while the phase was treated as a free parameter. In this way, we arrived at a likelihood map for each target, where for each period the model likelihood decreases with increasing planetary mass. 
We evaluated the differences in $\ln\mathcal{L}$ across the whole grids using the models with semi-amplitudes compatible with zero as null hypotheses and, consequently, the average of their corresponding likelihoods as the baseline.
From these differences, we derived a binary detection probability map using a threshold of $\Delta\ln\mathcal{L} = 5$ with respect to the null hypothesis. Given the implication that in regions with differences in log-likelihood below that threshold an additional signal cannot be ruled out and, therefore, could be hidden in the data, the corresponding planets could not be detected. Consequently, analogously to \cite{Tuomi2014}, we assumed that signals from planets within the complementary area of the mass-orbit plane, with differences in log-likelihood above the threshold, would likely be detected if they were apparent in the given time series. Therefore, we applied a detection probability of 100\,\% for $\Delta\ln\mathcal{L} > 5$ and 0\,\% for $\Delta\ln\mathcal{L} < 5$. 
This threshold corresponds to a probability ratio between two distinct models of about 150, and has been used in RV data analyses in the past~\citep[e.g.,][]{Feroz2011,Gregory2011, Tuomi2012}.
The individual binary maps were then averaged across the entire grid and the stellar sample in order to arrive at the final sensitivity map, which is illustrated in \autoref{fig:sensmaps_loL_and_comp}.
The alternative occurrence rates based on these detection probabilities were calculated thereafter following the same procedure as outlined in \autoref{subsubsec: occrates}, and the results are listed also in \autoref{tab:occurrence_all_merged}. 

Although the absolute numbers differ slightly from those derived using the sensitivity map based on injection-and-retrieval, they agree within their uncertainties and, moreover, they also show the same dependencies with orbital periods and planetary masses. All of this is plausible, as the trends over different period and mass intervals are mostly governed by our actual sample of detected planets, while the magnitude of the absolute numbers depends highly on the sensitivity map used to account for the detection bias. The differences in sensitivity between the two approaches are depicted as differences of detection probabilities $\mathcal{P}_\text{inj.-retr.}-\mathcal{P}_\text{lik.}$ in \autoref{fig:sensmaps_loL_and_comp}. 

From this comparison, a general trend is apparent for orbital periods above around 5\,d. For a given period, the detection probabilities based on the likelihood maps tend to be a bit lower for smaller planetary masses but exceed those based on injection-and-retrieval for higher masses. For shorter orbital periods ($P_{\rm pl} <$ 5\,d) the alternatively derived probabilities are typically higher. The mean of the differences over the entire grid is --2.9\,\%. The comparably lower detection probabilities from the likelihood maps at low planetary masses but wider orbits lead to increased occurrence rates in this region of the grid, whereas the rates are slightly lower than those from the injection-and-retrieval maps anywhere else. This trend is also confirmed by comparison of the probabilities averaged along the paths of the error limits, as can be seen in Figs. \ref{fig:planet_occurrence} and \ref{fig:sensmaps_loL_and_comp}. 

We stress the fact that, although the final numbers differ slightly, the results are still compatible, as both methods allow for some margin. Therefore, it is difficult to determine which method is superior.
Both methods involve choosing a somewhat arbitrary threshold as detection criteria. 
While for the injection-and-retrieval method one needs to choose a threshold for the FAP of signals being counted as retrieved, one is left with an arbitrary choice of a difference in the $\Delta\ln\mathcal{L}$ distribution at which one assumes a model not compatible with the null hypothesis any more. The limit of $\Delta\ln\mathcal{L} = 5$ that we chose translates into a probability ratio of $1/150\,\%=0.67\,\%$ (see above), which is comparable to the 1\% that we used in the injection-and-retrieval approach. 
Given these considerations, we believe that the compatibility of the resulting occurrence rates from the two methods underlines their robustness, in general. Still, one should keep in mind the margins given by the choice of the different applied methods when comparing the occurrence rates of different surveys and studies.

\subsection{Comparison to planet formation theory}

A comprehensive synthesis of planet formation around M dwarfs in the standard core accretion scenario was presented by \citet{Burn2021}. 
They used the model of \citet{Emsenhuber2021}, which integrates the growth by solid (in the form of planetesimals) and gas accretion, orbital migration, as well as N-body interactions of 50 concurrently growing protoplanets. At the start of the simulations, small planetesimals and the largest seed protoplanets are assumed to have formed. The initial disk properties are informed by Class I disk measurements and scaled to lower stellar masses.
Based on this model, \citet{Burn2021} found occurrence rates of Earth-like planets of order unity in agreement with our results presented here. 
Giant planets, however, as reported by \citet{Morales2019}, challenge planet formation models, as such ones cannot form around late M dwarfs under the standard assumptions~\citep{Schlecker2022}: The required planetesimal densities are not reached at orbital distances with sufficiently short growth timescales~\citep{Schlecker2021b}.
This conundrum might be solved by introducing disk structures acting as traps for migration~\citep[e.g.,][]{Hasegawa2011a}.

\begin{figure*}
    \centering
    \includegraphics[trim=0 0 0 0,clip,height=0.34\linewidth]{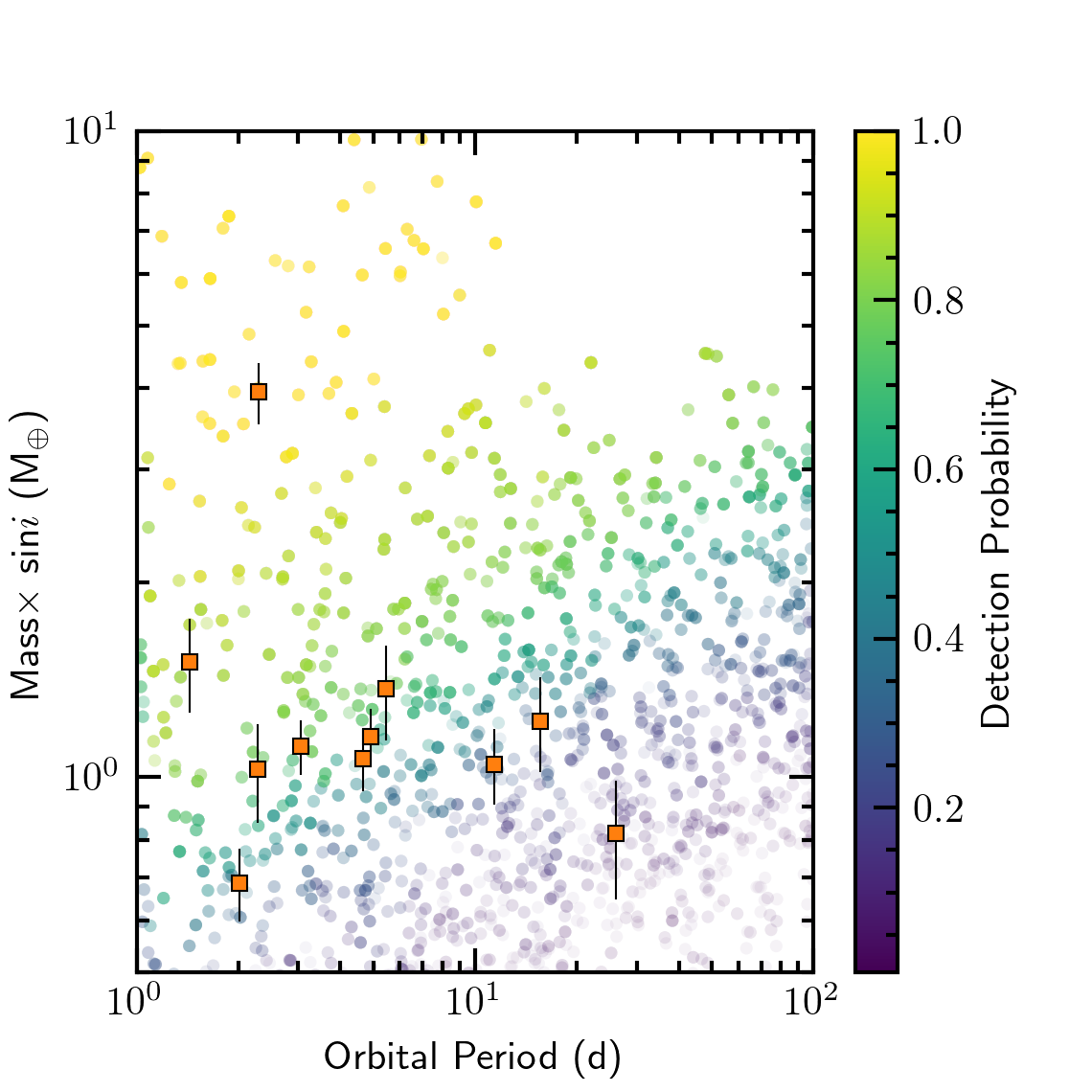}
    \includegraphics[trim=0.0cm 0 0.75cm 0,clip,height=0.34\linewidth]{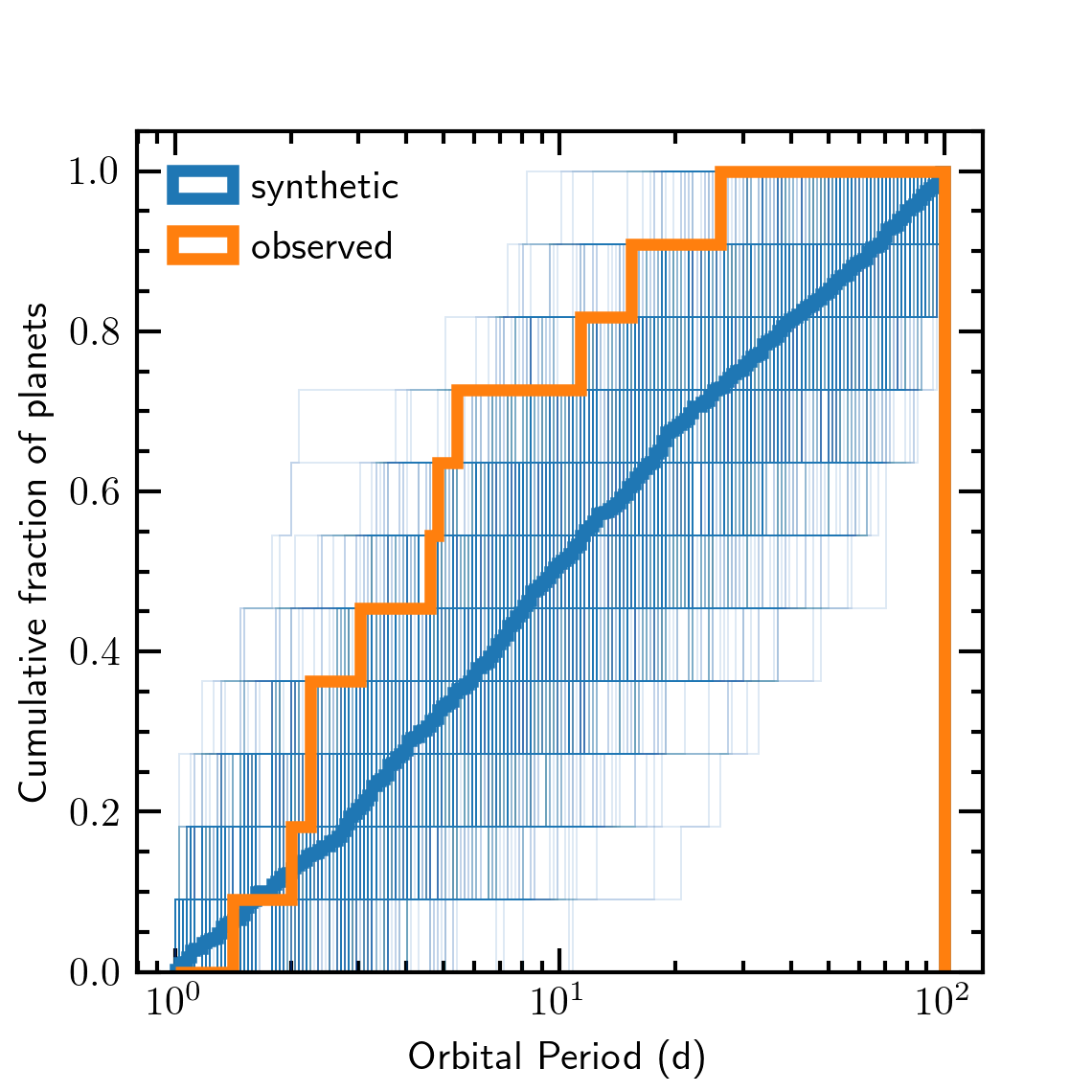}
    \includegraphics[trim=1.2cm 0 0 0,clip,height=0.34\linewidth]{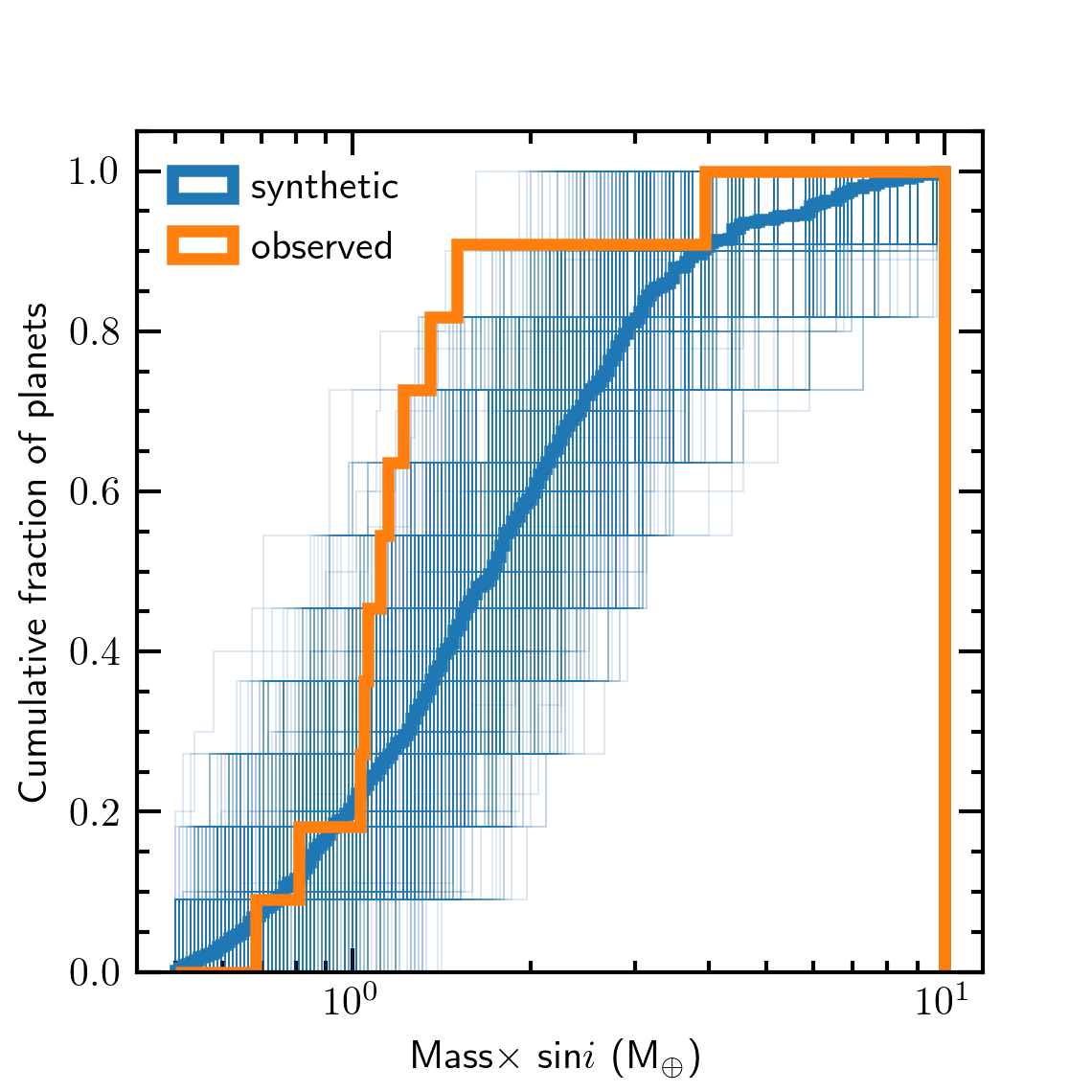}
    \caption{Planet detection statistics of observed planets compared with mock detections based on the synthetic planet population by \citet{Burn2021}. Included are all planets with orbital periods of 1--100\,d and with minimum masses of 0.5--10\,M$_\oplus$. The left panel shows the $M_{\rm pl} \sin i$ of the observed (orange squares) and synthetic (circles) planets against orbital periods. The transparency and color of the synthetic data are scaled by detection probability. The cumulative distribution of the observed and synthetic $P_{\rm pl}$ (middle) and $M_{\rm pl} \sin i$ (right) is shown. For the synthetic data, we show with transparent lines 1000 random draws of 11 planets to visualize the expected spread.}
    \label{fig:planet_occurrence_theory}
\end{figure*}

Under the standard assumption that accretion of planetesimals and N-body interactions between low-mass planets leading to giant impacts dominate in the inner disk region, we compared the model predictions with our determined occurrence rates. 
As in \citet{Schlecker2022}, we randomly drew inclinations for the systems synthesized by \citet{Burn2021} and created 1000 mock observations of 11 synthetic planets per draw. The chance of drawing each planet was weighted by the $M_{\rm pl} \sin i$ and orbital period-dependent detection sensitivities mentioned in \autoref{subsec:Det_completeness} and shown in \autoref{fig:planet_occurrence}.

The resulting distribution of synthetic planetary $M_{\rm pl} \sin i$ and orbital periods was then compared to the known small planets detected using CARMENES data only, which are listed in \autoref{table:planets}. The results shown in \autoref{fig:planet_occurrence_theory} indicate some significant differences: The hypothesis that $P_{\rm pl}$ and $M_{\rm pl} \sin i$ are individually drawn from the same distribution is excluded using a 1D Kolmogorov-Smirnov test at moderate $p$ values of 0.068 and 0.008, respectively. The observed planets orbit their host stars at orbital periods shorter than predicted, and their masses are smaller. In fact, due to the dependency of the detection sensitivity on orbital period and mass, both distributions can no longer be statistically distinguished if synthetic planetary masses are reduced by a factor of two ($p$ values of 0.5 and 0.3 for $P_{\rm pl}$ and $M_{\rm pl} \sin i$, respectively).

As planet mass is the most fundamental and constraining property for their formation, a disagreement about the population of small planets demands an explanation. While a thorough model iteration with different parameters is out of the scope of this work, we used analytical mass scales derived by \citet{Emsenhuber2023} to assess what changes in initial conditions would lead to consistent planetary masses. Assuming that the planets did not migrate over large separations, the relevant mass scale is the Goldreich mass

\begin{equation}
M_{\mathrm{Gold}} = 16 a^3  \Sigma_\mathrm{P}^{3/2} \left(\frac{2\pi^7 a^3 \rho_\mathrm{P}}{3 M_\star^3}\right)^{1/4}\,,
\label{eq:mgold}
\end{equation}

\noindent where $a$ is the semi-major axis, $\Sigma_\mathrm{P}$ is the surface density of planetesimals, and $\rho_\mathrm{P}$ is the bulk density of the rocky planets and planetary embryos. A reduction of the amount of planetesimals by 37\,\% is sufficient to decrease the expected mass of the planets by 50\,\% and, therefore, to the range of the observed planet mass distribution. For solar-mass stars, a comparison between the synthetic data from the same planet formation model and the HARPS survey results \citep{Mayor2011} shows no evidence that such a reduction is needed \citep{Emsenhuber_2024_subm}. Moreover, the planetary mass range can be probed to lower values than what HARPS can achieve around solar-type stars. Combined, those results hint at a steeper-than-linear dependency of the available mass in planet-building material with stellar mass.

Future studies should further explore whether a possible lower planetesimal surface density for the lowest-mass stars can be explained by planetesimal formation models. The results from \citet{Lenz2019} indicate even higher surface densities than those assumed by \citet{Burn2021} due to the predicted steeper slope of the radial surface density of about --2, in contrast to the adopted --1.5. However, \citet{Burn2021} assumed globally complete conversion of the dust mass to planetesimals, which is perhaps overly optimistic. Moreover, the used dust masses were observed by \citet{Tychoniec2018} in the Perseus association, which seems to be one of the star-forming regions with more massive disks as opposed to, for example, Ophiuchus (\citealp{Williams2019}, see also the discussion by \citealp{Tobin2020}). Furthermore, if preferential locations exist for planetesimal formation, such as the water ice line \citep{Drazkowska2017}, a different planet formation channel is required for close-in planets.

An alternative scenario to explain the differences between the observed and predicted planetary mass and orbit distributions is migration from the exterior of the snowline, where different mass scales can apply, namely the typical mass at which type-I migration dominates over the relevant solid accretion mechanism ($M_{\rm pl} \approx 5 M_\oplus$.). If this is the major origin of the planets, they should contain large amounts of H$_2$O and other volatile species, which are expected to be retained during photoevaporation \citep{Lopez2017,Burn2024}. So far, transiting planets in the mass range of the low-mass planets in \autoref{tab:planetparams} are consistent with being rocky \citep{Luque_2022}.

Another possibility is that the growth of planets is dominated by pebble accretion. In this case, the pebble isolation mass, which depends on the uncertain temperature and viscosity of the disk, can leave a distinct imprint on the mass distribution \citep{Brugger2020} similar to what is observed in Fig. \ref{fig:planet_occurrence_theory}. However, the work of \citet{Liu2019} focusing on late M dwarfs found lower planetary masses than found here (while the recent work of \citealp{Venturini2024} found rocky planets with masses up to 3\,$M_\oplus$). Therefore, subsequent giant impacts, excluded by \citet{Liu2019}, would be required between the planets at the pebble isolation mass, which would re-introduce scatter in the planetary masses.

Finally, an additional effect omitted in the simulations is the loss of material to debris during giant impacts. 
Although the simulations by \citet{Burn2021} assumed perfect merging, about a factor of two in mass is lost due to fragmentation in realistic giant impacts~\citep{Emsenhuber2020}.
When this mass loss is taken into account, the outcome agrees remarkably well with our results and provides an additional avenue to explain the population of small planets around very late M~dwarfs, in addition to reduced planetesimal surface densities and dry pebble accretion with subsequent migration.

\section{Summary} \label{sec:conclusions}

We studied planet occurrence rates in a sample of 15 late-type M dwarfs observed with the CARMENES spectrograph.
All the stars are low-mass ($M \lesssim$ 0.16\,M$_\odot$), relatively bright ($J <$ 10\,mag), slowly-rotating ($v \sin{i} <$ 2\,km\,s$^{-1}$), and weakly-active (pEW(H$\alpha$) $>$ --1.5\,\AA).  
We used available photometric data combined with time series of different spectral activity indicators from CARMENES to determine the rotation period for one of the targets without a previous measurement.

In order to minimize possible detection biases in our final occurrence rates, we performed an injection-and-retrieval analysis on the RV data of our targets to determine the detection sensitivities with respect to all data sets at hand. In the process of reevaluating the RV time series and identifying all significant periodic signals, we confirmed ten known planets around five M dwarfs and discovered four new planets around three stars:
G~268--110\,b, with a minimum mass of $M=\MassJZero$ on a short orbit of 
$P \approx$ 1.43\,d around its host,
G~261--6\,b, a $\MassJOne{}$ planet with a period of 
$P \approx$ 5.45\,d,
and two companions around G~192--15. One of them is an Earth-mass planet with $1.03 \pm 0.18$\,M$_\oplus$ at 
$P \approx$ 2.27\,d and the other is a $14.3 \pm 1.6$\,M$_\oplus$ planet on a wide and eccentric orbit with 
$P \approx$ 1220\,d and $e = 0.68 \pm 0.07$.

The final planet occurrences were determined by Monte Carlo simulations within predefined orbital period and minimum mass ranges. We took into account the overall detection probabilities averaged over the individual targets' detection sensitivity maps in order to account for false statistics from missing detections. 
We found an apparent trend in planet occurrence as a function of minimum planet masses. The rates decrease substantially from $\overline{n}_\text{pl} = 0.88^{+0.36}_{-0.28}$ for planetary masses between 0.5\,M$_\oplus$ and 3\,M$_\oplus$ to $\overline{n}_\text{pl} = 0.11^{+0.11}_{-0.06}$ for planetary masses between 3\,M$_\oplus$ and 10\,M$_\oplus$ in the orbital period regime from $P=1$\,d to $P=10$\,d, and from $\overline{n}_\text{pl} = 0.92^{+0.56}_{-0.39}$ to $\overline{n}_\text{pl} = 0.06^{+0.09}_{-0.04}$ respectively for orbital periods between $P=10$\,d and $P=100$\,d. As our results do not show any significant dependency on orbital periods, we could not confirm the general trend of increasing occurrence rates when moving from wider to shorter orbits, as indicated by \cite{Sabotta2021} in a larger mass regime of M dwarfs ($M$ = 0.095--0.34\,M$_\odot$). However, in comparison to previous studies \citep[e.g.,][]{Bonfils_2013_sample,Hardegree-Ullman2019,Pinamonti_2022}, we showed that stars of later spectral types tend to host a larger number of small planets on shorter orbits. 
This trend is apparent among M dwarfs, as for our sample of targets with the lowest masses we found at least twice as many planets on shorter orbits than indicated by the occurrence rates from \cite{Ribas2023}, who analyzed the entire CARMENES sample, back then covering the mass range $M$ = 0.095--0.677\,M$_\odot$.

Since occurrence rates are highly affected by possible detection biases, a proper estimation of the apparent detection limits is crucial. Although for our analysis we relied on an injection-and-retrieval analysis, other methods, such as posterior sampling have been applied for this purpose in the past \citep{Tuomi2014, Pinamonti_2022}. Therefore, we tested another method, based purely on likelihood maps. While the resulting sensitivity maps differ from our initial approach, the final occurrence rates are robust against the choice of method and agree within their uncertainties. The rather high uncertainties in our rate estimations presumably stem mainly from low-number statistics, as our final stellar sample consists of only 15 targets with 11 known planets that satisfy our selection criteria. All except one of these planets have minimum masses around or below 1.5\,M$_\oplus$, and in particular below the minimum mass of water worlds, so that their composition is probably rocky \citep{Luque_2022}. However, with only one planet in the highest mass bin, the rates for the mass bin above 3\,M$_\oplus$ should be handled with caution.

To put our results in context, we compared the currently known planets around the stars within our stellar sample with predictions from state-of-the-art planet formation models. We simulated mock observations of planets drawn from a synthetic planet population around M dwarfs based on the standard core accretion scenario \citep{Burn2021}, and weighted them by our estimated detection probabilities. The observed orbital periods are shorter and the minimum masses are smaller than predicted. In the framework of standard core accretion, such a deviation could in principle be explained by a reduced planetesimal surface density in models. While such a correction was not needed based on previous occurrence rate studies, we probed here the very low-mass star regime. Therefore, the dependency of available mass in planet building blocks on stellar mass needs to be further investigated. Apart from that, the observed lower masses could also be explained by alternative formation scenarios such as accretion of dry pebbles within the snowline or mass loss during giant impacts.

Altogether, our results and the discussion show the importance of differentiation in stellar masses when discussing planet occurrences and, ultimately, planet formation processes. Once the CARMENES survey of M dwarfs is completed and at least 50 RV epochs have been obtained for all targets, we will conclude our analysis of occurrence rates on the entire sample. 

\begin{acknowledgements}
  We thank the anonymous referee for a very quick and
  constructive report.

  This publication was based on observations collected under the CARMENES Legacy+ project.
  
  CARMENES is an instrument at the Centro Astron\'omico Hispano en Andaluc\'ia (CAHA) at Calar Alto (Almer\'{\i}a, Spain), operated jointly by the Junta de Andaluc\'ia and the Instituto de Astrof\'isica de Andaluc\'ia (CSIC).
    
  CARMENES was funded by the Max-Planck-Gesellschaft (MPG), 
  the Consejo Superior de Investigaciones Cient\'{\i}ficas (CSIC),
  the Ministerio de Econom\'ia y Competitividad (MINECO) and the European Regional Development Fund (ERDF) through projects FICTS-2011-02, ICTS-2017-07-CAHA-4, and CAHA16-CE-3978, 
  and the members of the CARMENES Consortium 
  (Max-Planck-Institut f\"ur Astronomie,
  Instituto de Astrof\'{\i}sica de Andaluc\'{\i}a,
  Landessternwarte K\"onigstuhl,
  Institut de Ci\`encies de l'Espai,
  Institut f\"ur Astrophysik G\"ottingen,
  Universidad Complutense de Madrid,
  Th\"uringer Landessternwarte Tautenburg,
  Instituto de Astrof\'{\i}sica de Canarias,
  Hamburger Sternwarte,
  Centro de Astrobiolog\'{\i}a and
  Centro Astron\'omico Hispano-Alem\'an), 
  with additional contributions by the MINECO, 
  the Deutsche Forschungsgemeinschaft (DFG) through the Major Research Instrumentation Programme and Research Unit FOR2544 ``Blue Planets around Red Stars'' (RE 2694/8-1), 
  the Klaus Tschira Stiftung, 
  the states of Baden-W\"urttemberg and Niedersachsen, 
  and by the Junta de Andaluc\'{\i}a.
  
  We used data from the CARMENES data archive at CAB (CSIC-INTA).
  
  We acknowledge financial support from the Agencia Estatal de Investigaci\'on (AEI/10.13039/501100011033) of the Ministerio de Ciencia e Innovaci\'on and the ERDF ``A way of making Europe'' through projects 
  PID2022-137241NB-C4[1:4],	
  PID2021-125627OB-C31,		
and the Centre of Excellence ``Severo Ochoa'' and ``Mar\'ia de Maeztu'' awards to the Instituto de Astrof\'isica de Canarias (CEX2019-000920-S), Instituto de Astrof\'isica de Andaluc\'ia (CEX2021-001131-S) and Institut de Ci\`encies de l'Espai (CEX2020-001058-M).

  This work was also funded by the Generalitat de Catalunya/CERCA programme, 
  the DFG under Germany’s Excellence Strategy EXC 2181/1-390900948 Exploratory project EP~8.4 (Heidelberg STRUCTURES Excellence Cluster),
  and the Bulgarian National Science Fund (FNI) program ``VIHREN-2021'' project No. KP-06-DV/5.

  The results reported herein benefitted from collaborations and/or information exchange within NASA’s Nexus for Exoplanet System Science (NExSS) research coordination network sponsored by NASA’s Science Mission Directorate under Agreement No. 80NSSC21K0593 for the program ``Alien Earths''.

\end{acknowledgements}

\bibliographystyle{aa} 
\bibliography{bibtex.bib} 

\begin{appendix}
\onecolumn
\section{Rotational period of G~109--35}\label{app:rot_period}
For the determination of the stellar rotational period of G~109--35 (Karmn J06594+193) we used all available photometric data and spectroscopic activity indicators. The activity indicators are measured from CARMENES spectra and are available as a time series of 30 data points corresponding to our RV measurements collected over four years. As two observations did not provide information on drift corrections during the night, the corresponding precise RVs were discarded, while the determination of the activity indicators is still valid. 

\begin{figure*}[h!]
    \centering
    \includegraphics[width=\linewidth]{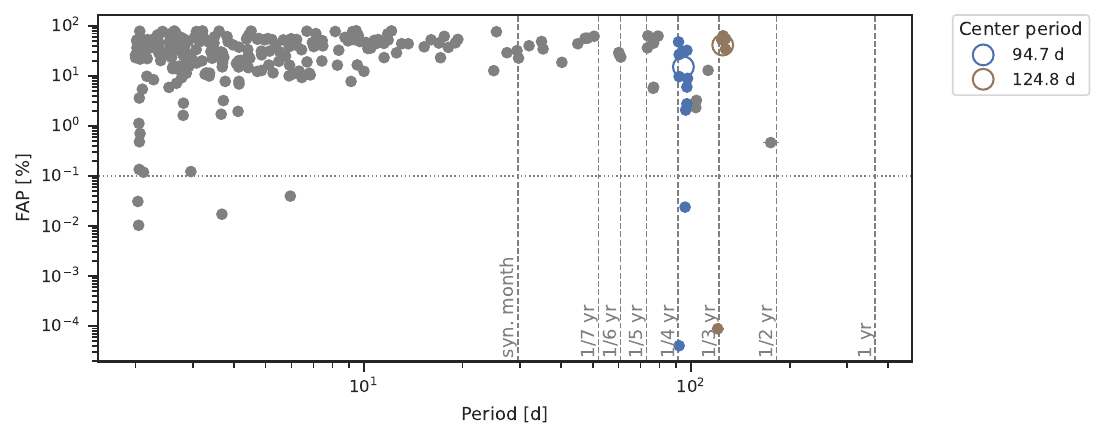}
    \caption{Diagram of peaks appearing in the GLS periodograms of the activity indicators accessible with CARMENES for G~109--35 (J06594+193). 
    Big open circles at 94.7\,d (in blue) and 124.8\,d (brown) mark the center periods of the two main {\tt DBSCAN} clusters.
    Vertical dashed lines indicate one year and its higher harmonics, while the horizontal dotted line indicates FAP = 0.1\,\%. 
    }
    \label{fig:activity_clusters_J06594+193}
\end{figure*}

To search for common periods in the GLS periodograms of the indicators, we applied the {\tt DBSCAN} clustering algorithm as described in \autoref{subsec:G268-110}.
The result of this analysis is illustrated in \autoref{fig:activity_clusters_J06594+193}.
It revealed two clusters of signals at periods of around \SI{95}{\day} and \SI{125}{\day}, but with only three activity signals being significant at FAPs of $\num{4.0E-07}$--$\num{2.0E-04}$ and $\num{8.8E-07}$, respectively.
We used the positions of those signals as prior knowledge in our further analysis. 

We then analyzed photometric data from the All-Sky Automated Survey (ASAS\footnote{\url{http://www.astrouw.edu.pl/asas}}; \citealt{Pojmanski1997}) and the Northern Sky Variability Survey (NSVS\footnote{\url{https://skydot.lanl.gov/nsvs/}}; \citealt{Wozniak2004}) as compiled by \cite{DiezAlonso2019}. 
The ASAS data available to us cover around seven years between December 2002 and November 2009, and were taken by the $V$-band wide-field camera at the survey's station located in Las Campanas Observatory, Chile (ASAS-3), while the NSVS photometric measurements that we used were collected in Los \'Alamos, USA, 
between April 1999 and April 2000. An inspection of the corresponding GLS periodograms did not reveal any significant signals at the periods of interest, as already reported by \cite{DiezAlonso2019}. 
We therefore continued with Gaussian process (GP) regression fits to the combined photometric data using a double harmonic oscillator (dSHO) kernel. A description of the kernel and its application to photometric data using \texttt{juliet} 
was provided by \cite{Kossakowski2021}.
The default priors for the parameters of the kernel used in this modeling are listed in \autoref{tab:priors_GPs}. During a first step, we used a wide unconstrained prior for the period to look for promising signals, of $\mathcal{U}$(10,200)\,d. The posterior distribution shows evidence for periods between \SI{90}{\day} and \SI{130}{\day}, as well as some at a period of around half a year. Next, using our prior findings from the analysis of the activity indicators, we constrained the range of possible rotation periods to further sample the regions around the 
two clusters in \autoref{fig:activity_clusters_J06594+193}. The shape of the posterior of this model is bimodal with peaks at periods of \SI{101}{\day} and \SI{119}{\day}. Although the shorter period appears to be slightly more favored in this distribution, we still continued to test both periods with models, for which the priors in periods were further narrowed, in order to sample those peaks individually. The results of these models yielded rotational periods of $P_\text{rot,1} = 101.9\pm5.0$\,d and $P_\text{rot,2} = 119.0\pm7.2$\,d, respectively. Unfortunately, the difference between the likelihoods of the two models is marginal and, therefore, does not indicate which of them is favored by the given data. For this reason, we declared both periods as genuine candidates for the true rotation period but noted that activity appears to be slightly more significant at the shorter one, as well as that it is slightly more favored during sampling with priors allowing both of them. Eventually, we report a single value $P_\text{rot}$ =110$^{+16}_{-13}$\,d in \autoref{table:stellar_properties}, which is determined by averaging the two candidate periods. Their original errors were used to derive the final uncertainties.

\FloatBarrier
\clearpage

\section{Analyses on planet discoveries}
\subsection{Discovery of G~268--110 b}
\label{app:J01048-181}

\FloatBarrier

\begin{figure*}[h!]
    \centering
    \includegraphics[width=\textwidth]{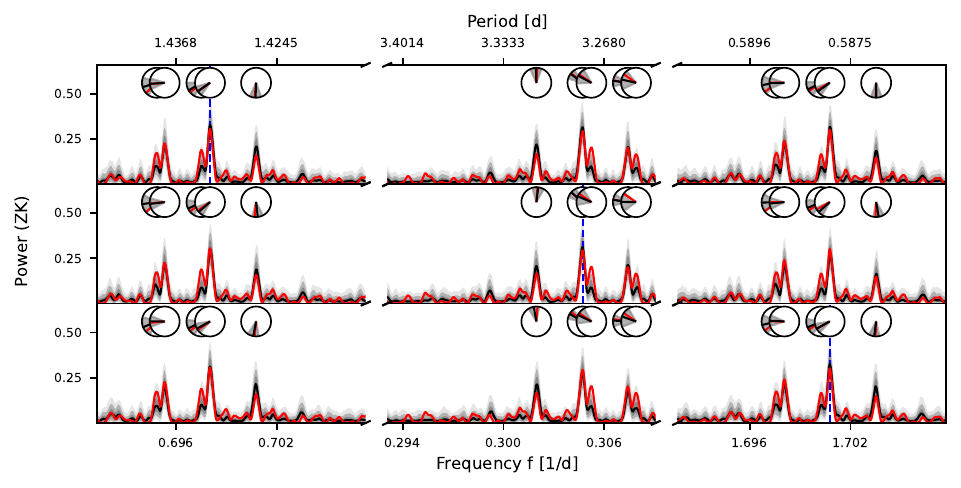}
    \caption{Alias test for \StarJZero{}. Each row represents the result of simulating a different underlying period: \textit{first row:} \SI{1.43}{\day}, \textit{second row:} \SI{3.28}{\day}, \textit{third row:} \SI{0.59}{\day} (each marked by the blue dashed line, respectively). The solid black line shows the median periodogram from 1000 simulations, with the interquartile ranges and the ranges of \SI{90}{\percent} and \SI{99}{\percent} denoted by the different gray-shaded areas. The observed periodogram is depicted by the red solid line. In addition to the periodograms, the resulting peak phases can also be compared in the circles above (the same colors as for the periodograms, but with the gray shade showing the standard deviation of the sampled peak phases).}
    \label{fig:aliasing_J01048-181}
\end{figure*}

\begin{figure*}[h!]
\setlength{\lineskip}{0pt}
    \centering
    \includegraphics[width=\textwidth, trim={0 0.73cm 0 0},clip]{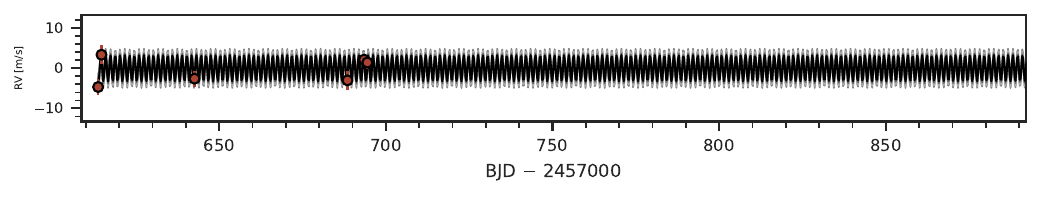}
    \includegraphics[width=\textwidth, trim={0 0.73cm 0 0},clip]{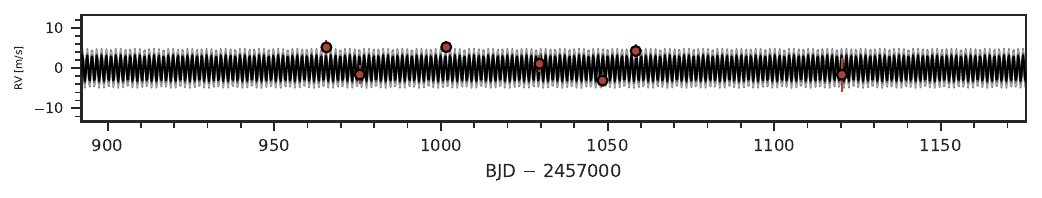}
    \includegraphics[width=\textwidth, trim={0 0.73cm 0 0},clip]{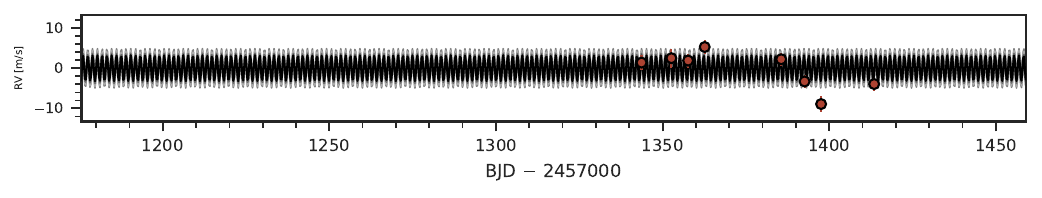}
    \includegraphics[width=\textwidth, trim={0 0.73cm 0 0},clip]{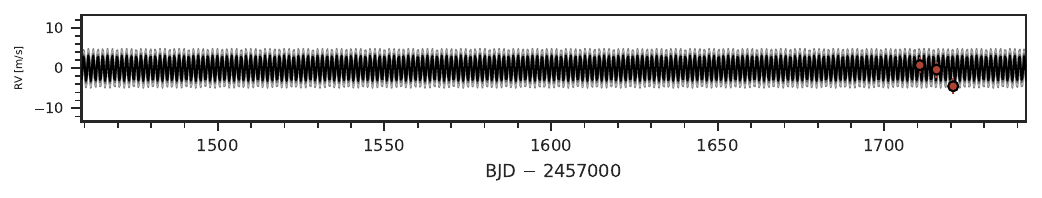}
    \includegraphics[width=\textwidth, trim={0 0.73cm 0 0},clip]{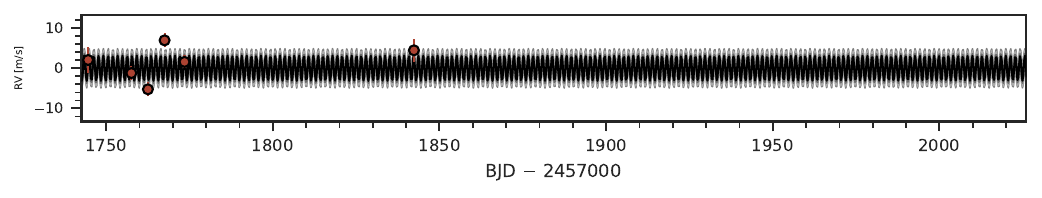}
    \includegraphics[width=\textwidth, trim={0 0.73cm 0 0},clip]{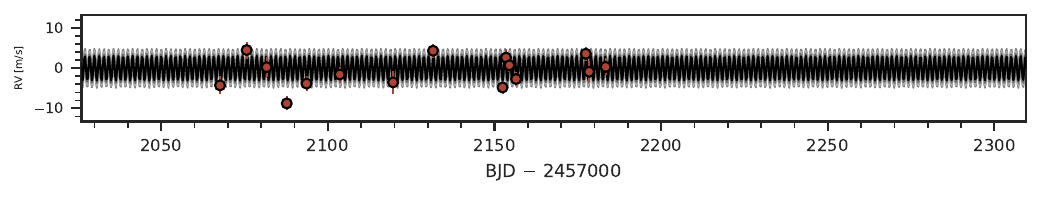}
    \includegraphics[width=\textwidth, trim={0 0.73cm 0 0},clip]{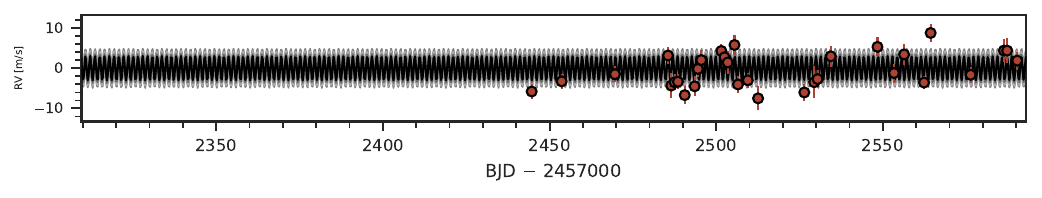}
    \includegraphics[width=\textwidth, trim={0 0.2cm 0 0},clip]{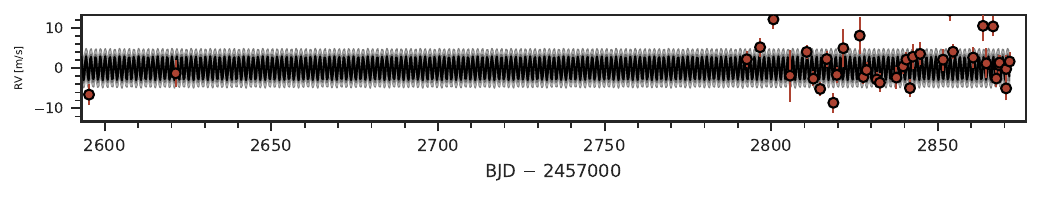}
    \caption{RVs over time for the best model ($=\text{1P}_\text{(1.4\,d-circ)}$) fitted to the CARMENES RVs of \StarJZero{} . The black lines show the model based on the parameters listed in \autoref{tab:planetparams}, and the gray shaded areas denote the \SI{68}{\percent}, \SI{95}{\percent,} and \SI{99}{\percent} confidence intervals, respectively. The instrumental offset of CARMENES was subtracted from the measurements and the model.}
    \label{fig:rvs_long_multipanel}
\end{figure*}

\begin{figure*}[h!]
    \centering
    \includegraphics[width=\textwidth]{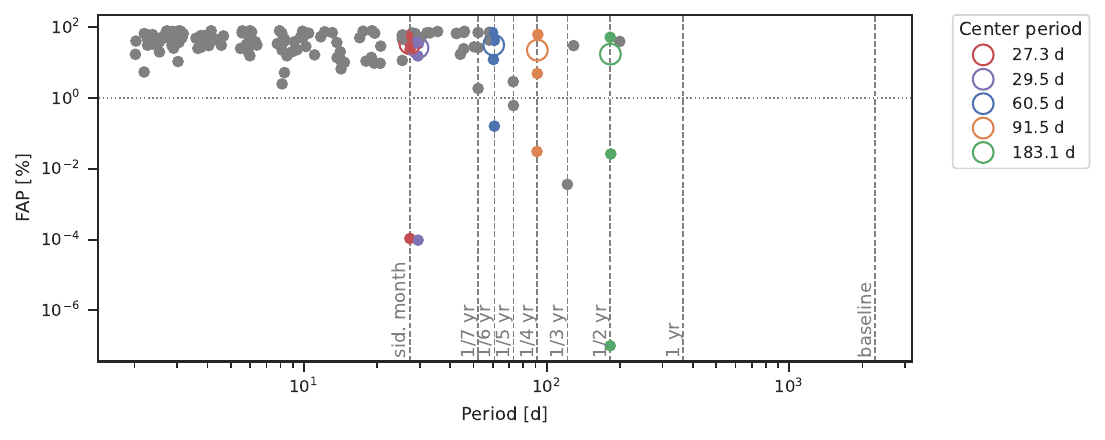} 
    \caption{Diagram of peaks appearing in the GLS periodograms of the activity indicators accessible with CARMENES for \StarJZero{}. For each activity indicator, the GLS periodogram was created and the ten highest-occurring peaks determined. If the FAP of the GLS peak is below \SI{80}{\percent}, it is written to a table. This list of peaks is used to run a DBSCAN clustering algorithm. A cluster is a group of peaks with at least 3 members, where the distance to the nearest neighbor is less than the resolution of the GLS.}
    \label{fig:activity_clusters_J01048-181}
\end{figure*}

\begin{figure*}[h!]
    \centering
    \includegraphics[width=\textwidth]{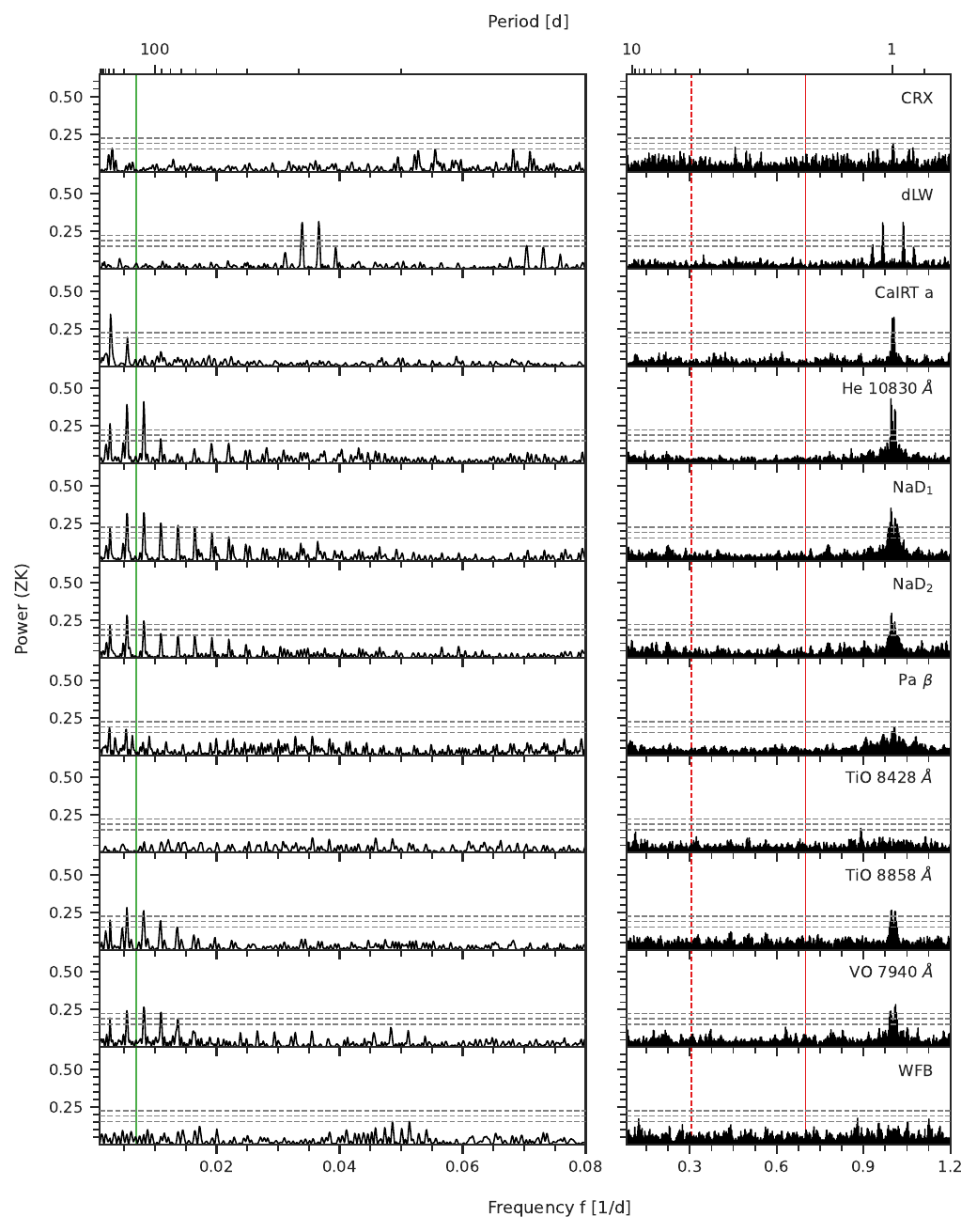}
    \caption{GLS periodograms of the activity indicators with signals of less then \SI{10}{\percent} FAP for \StarJZero{}. The period of the 1.4-day planet and the 3.3-day alias are highlighted by the red solid and dashed lines, respectively. The rotation period of \SI{143}{\day} determined by \cite{Newton2018} is marked by the green solid line.}
    \label{fig:activity_GLS_J01048-181_onecol}
\end{figure*}

\begin{table*}[h!]
\caption{Alternative fit results for the two alias periods of the 1.4-day planet signal.}
\label{tab:alt_planetparameters_J01048-181}
\centering
{\setlength{\extrarowheight}{4.5pt}
\begin{tabular}{cccc}
\hline \hline
Parameter & $P=\SI{0.5}{\day}$\tablefootmark{(a)} & $P=\SI{3.3}{\day}$\tablefootmark{(a)} & Units \\
\hline
$P_\text{b}$ & $\num{0.587972}^{+\num{1.2e-05}}_{-\num{1.3e-05}}$ & $\num{3.28169}^{+\num{0.00043}}_{-\num{0.00040}}$ & d \\
$t_{0,b}$ & $\num{2457613.578}^{+\num{0.038}}_{-\num{0.037}}$ & $\num{2457612.57}^{+\num{0.21}}_{-\num{0.24}}$ & d \\
$K_\text{b}$ & $\num{3.20}^{+\num{0.50}}_{-\num{0.51}}$ & $\num{3.17}^{+\num{0.50}}_{-\num{0.51}}$ & \si{\meter\per\second} \\
\noalign{\medskip}
$M_\text{p}\sin i$ & $\num{1.11}^{+\num{0.19}}_{-\num{0.19}}$ & $\num{1.95}^{+\num{0.33}}_{-\num{0.32}}$ & $M_\oplus$ \\
$a_\text{p}$ & $\num{0.00709}^{+\num{0.00016}}_{-\num{0.00016}}$ & $\num{0.0223}^{+\num{0.00048}}_{-\num{0.00050}}$ & \si{\astronomicalunit} \\
$T_\textnormal{eq, p}$\tablefootmark{({c})} & $\num{719.0}^{+\num{16.0}}_{-\num{16.0}}$ & $\num{405.2}^{+\num{8.7}}_{-\num{8.4}}$ & \si{\kelvin} \\
\hline
\end{tabular}}
\tablefoot{\tablefoottext{a}{Error bars denote the $68\%$ posterior credibility intervals.}}
\end{table*}

\clearpage

\subsection{Discovery of G~261--6 b}
\label{app:J19242+755}

\begin{figure*}[h!]
\centering
\includegraphics{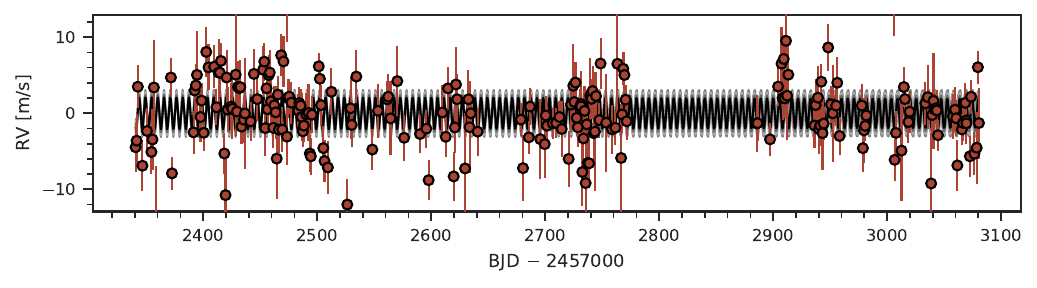}
\caption{Same as Fig.~\ref{fig:rvs_long_multipanel} but for the best model ($=\text{1P}_\text{(5\,d-circ)}$) fitted to the CARMENES RVs of \StarJOne{}.} 
\label{fig:rvs_over_time_J19242+755}
\end{figure*}

\begin{figure*}[h!]
    \centering
    \includegraphics[width=0.32\textwidth]{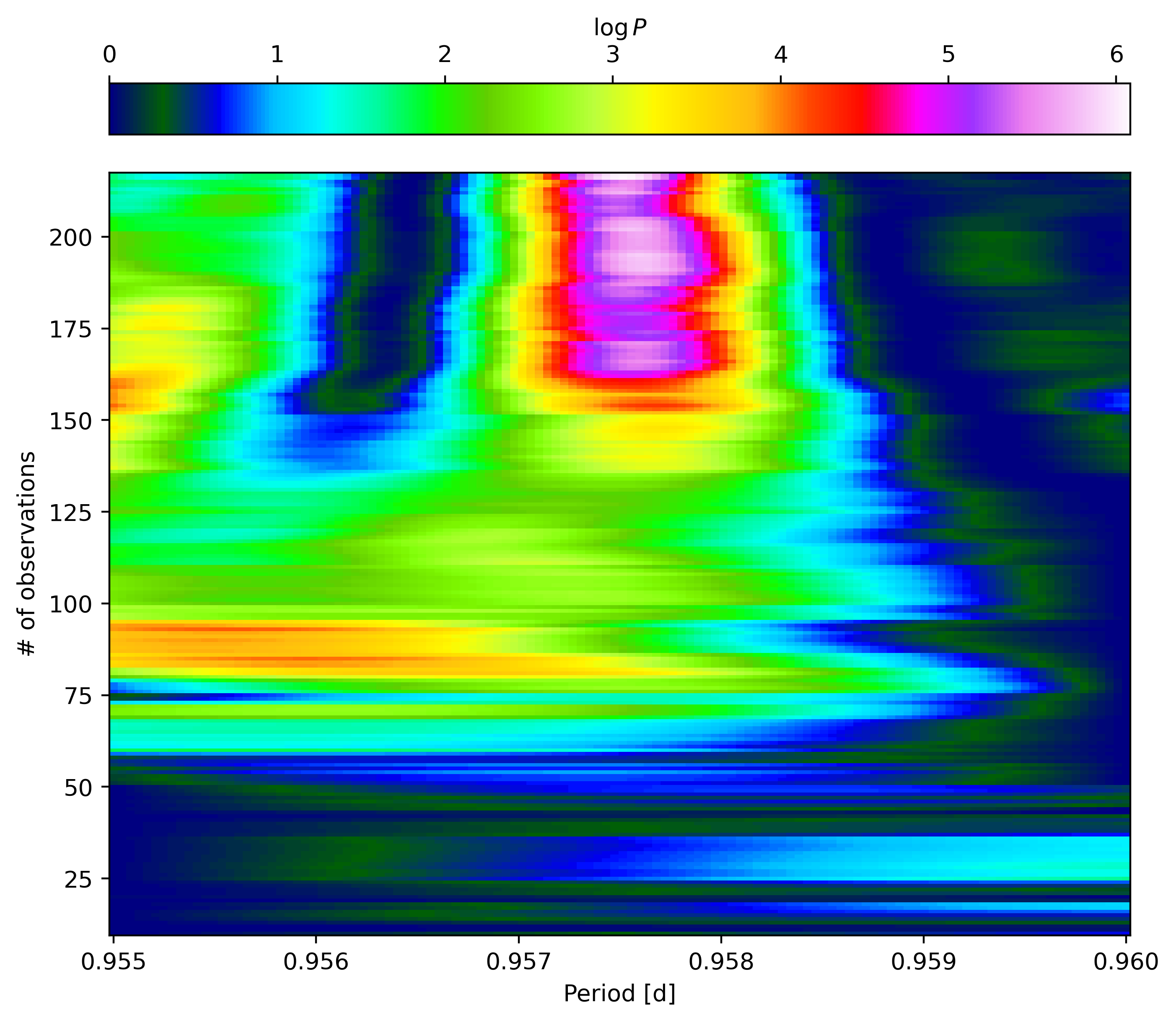} 
    \includegraphics[width=0.32\textwidth]{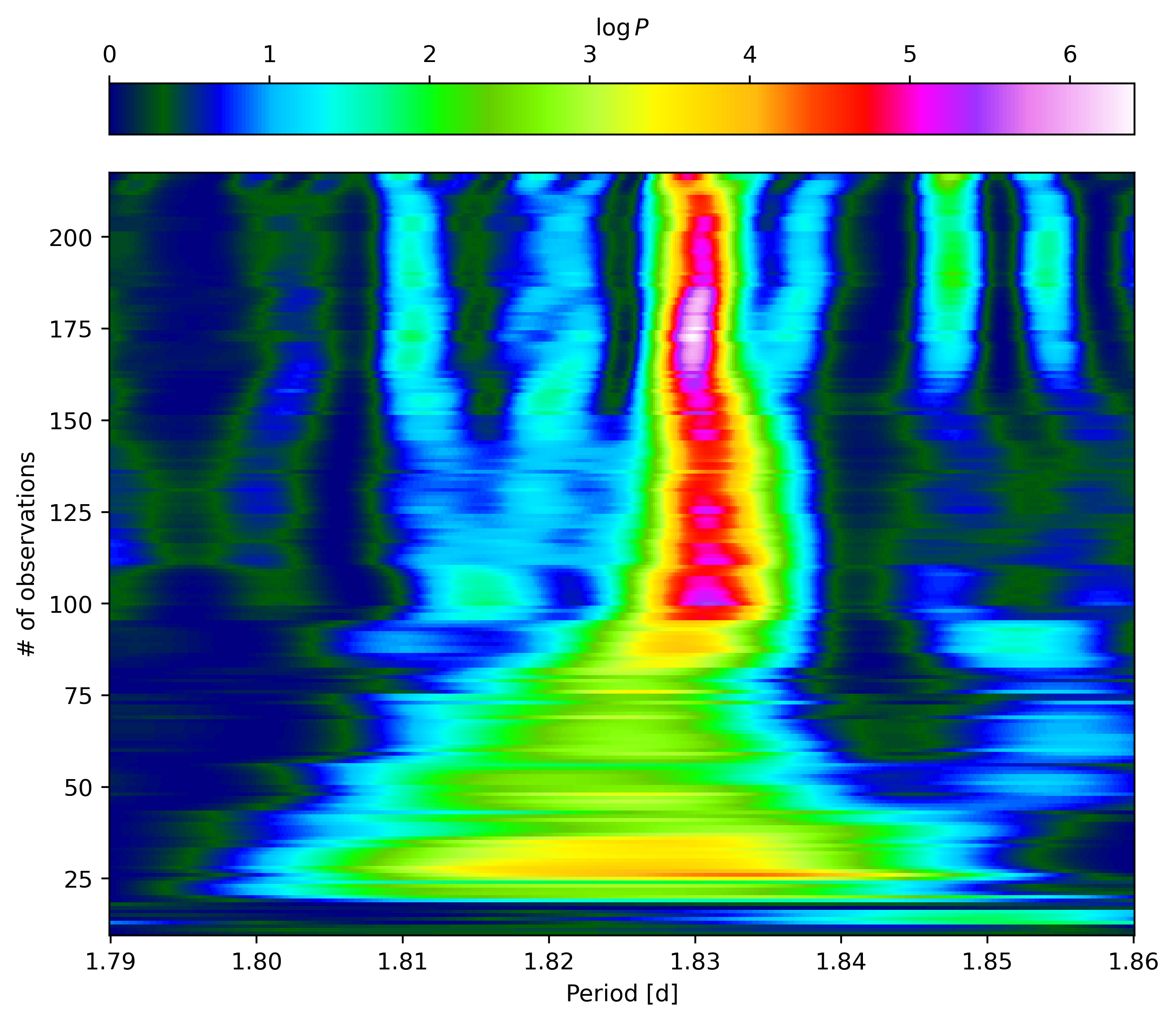}
    \includegraphics[width=0.32\textwidth]{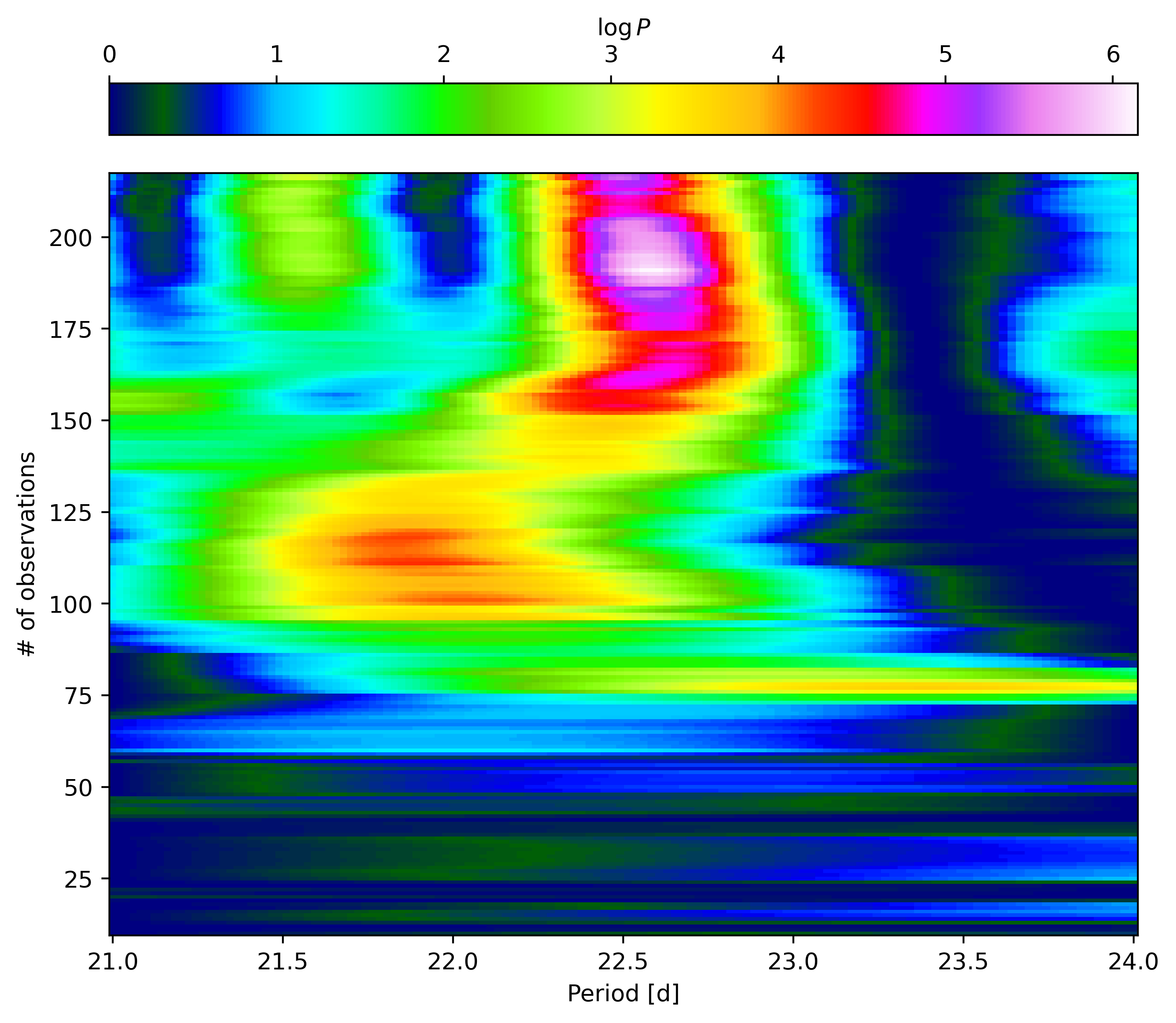} 
    \caption{Stacked Bayesian GLS periodograms of the three noteworthy signals with periods around 0.95\,d, 1.83\,d and 22.49\,d, detected in the residuals of the RVs after the planetary signal is removed.}
    \label{fig:sBGLS_J19242+755}
\end{figure*}

\begin{figure*}[h!]
    \centering
    \includegraphics[width=\textwidth]{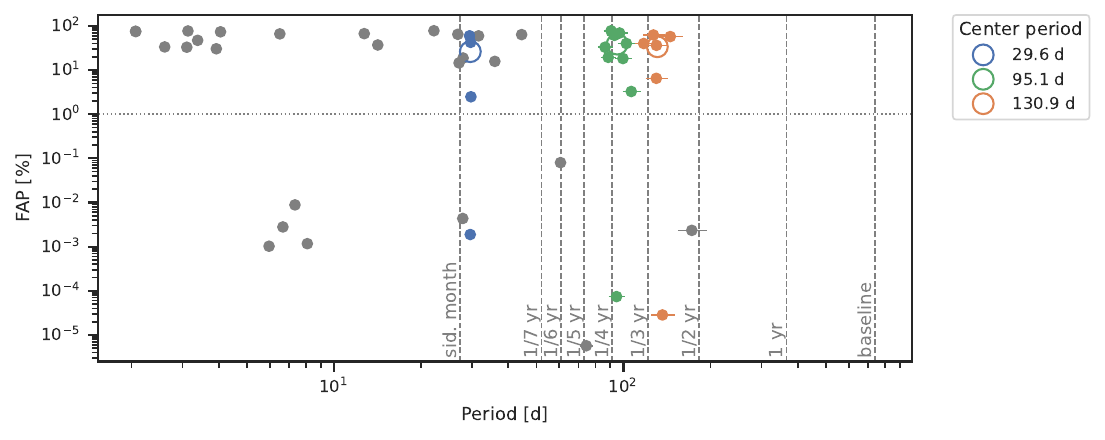} 
    \caption{Diagram of peaks appearing in the GLS periodograms of the activity indicators accessible with CARMENES for \StarJOne{}. For each activity indicator, the GLS periodogram was created and the ten highest-occurring peaks determined. If the FAP of the GLS peak is below \SI{80}{\percent}, it is written to a table. This list of peaks is used to run a DBSCAN clustering algorithm. A cluster is a group of peaks with at least 3 members, where the distance to the nearest neighbor is less than the resolution of the GLS.}
    \label{fig:activity_clusters_J19242+755}
\end{figure*}

\begin{figure*}[h!]
    \centering
    \includegraphics[width=\textwidth]{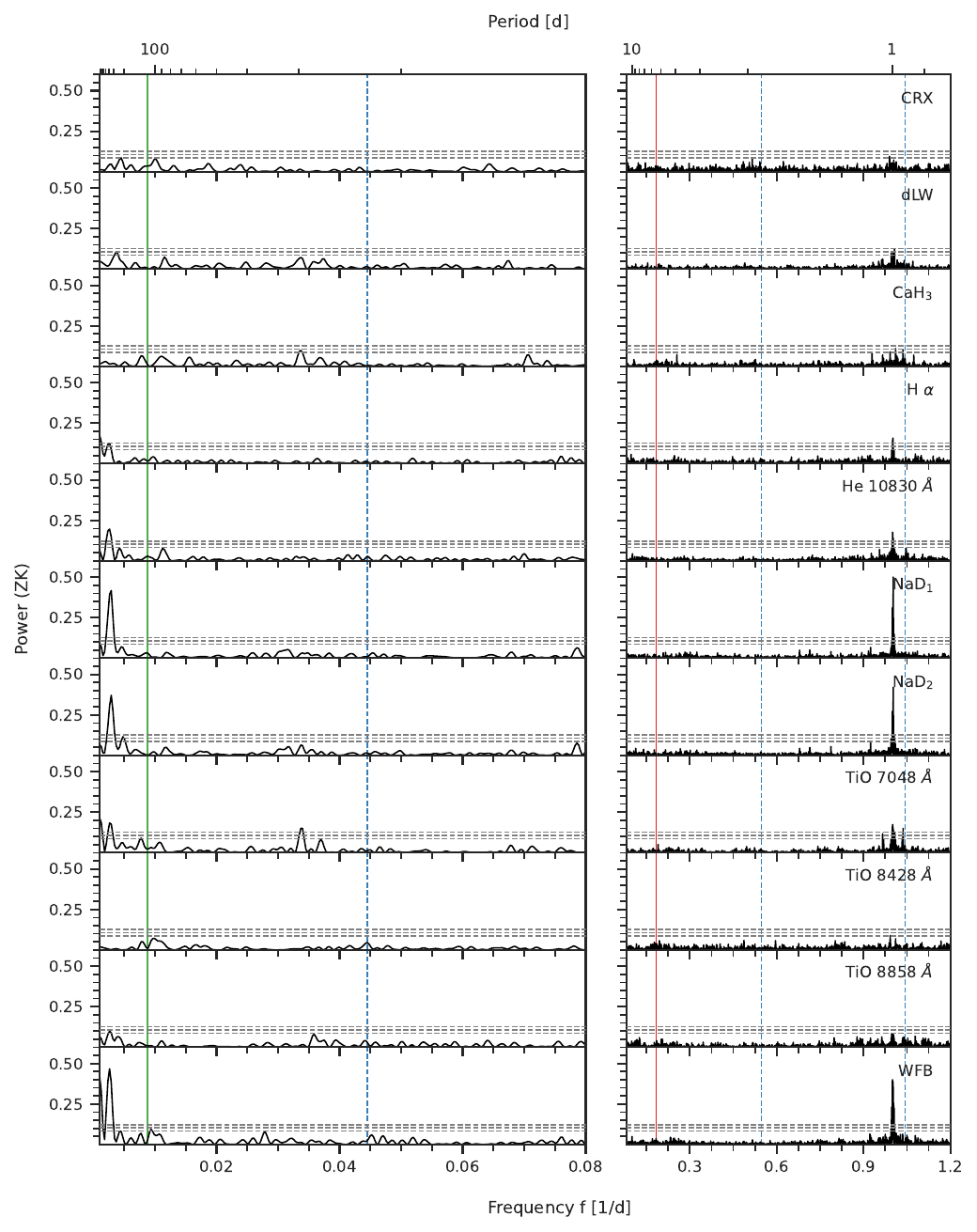}
    \caption{Same as Fig.~\ref{fig:activity_GLS_J01048-181_onecol} but for the activity indicators with significant signals for \StarJOne{}. The period of the 5.5-day planet is highlighted by the red solid line, while the locations of the possible candidate signals in the residuals at \SI{0.95}{\day}, \SI{1.83}{\day} and \SI{22.49}{\day} are indicated by the blue dashed lines, respectively. The rotation period of \SI{114}{\day} determined by \cite{Irwin2011} is marked by the green solid line.}
    \label{fig:activity_GLS_J19242+755}
\end{figure*}

\clearpage
\subsection{Discovery of G~192--15 b and G~192--15 c}
\label{app:J06024}
\begin{figure*}[h!]
    \centering
    \includegraphics[width=0.32\textwidth]{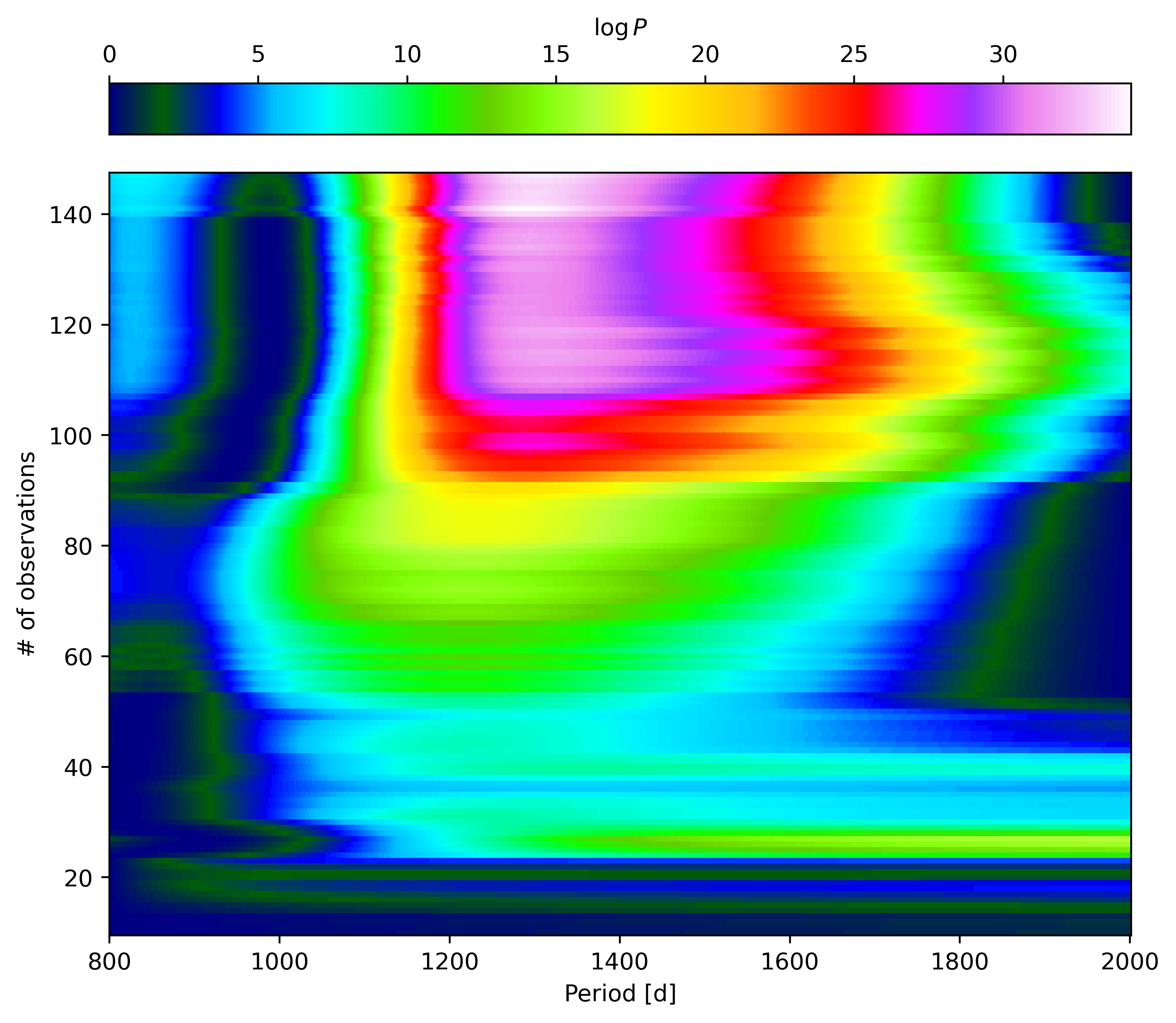} 
    \includegraphics[width=0.32\textwidth]{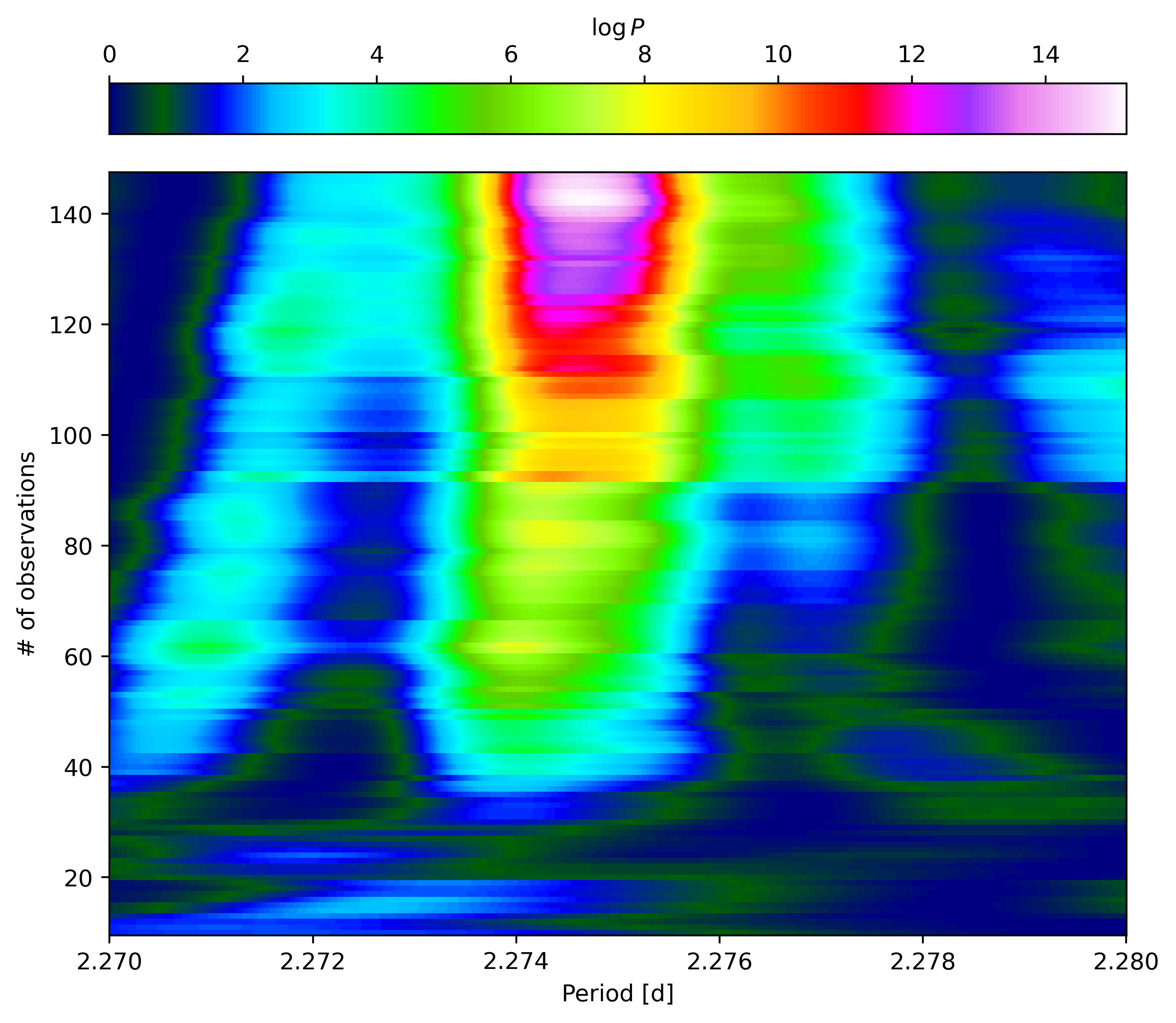}
    \includegraphics[width=0.32\textwidth]{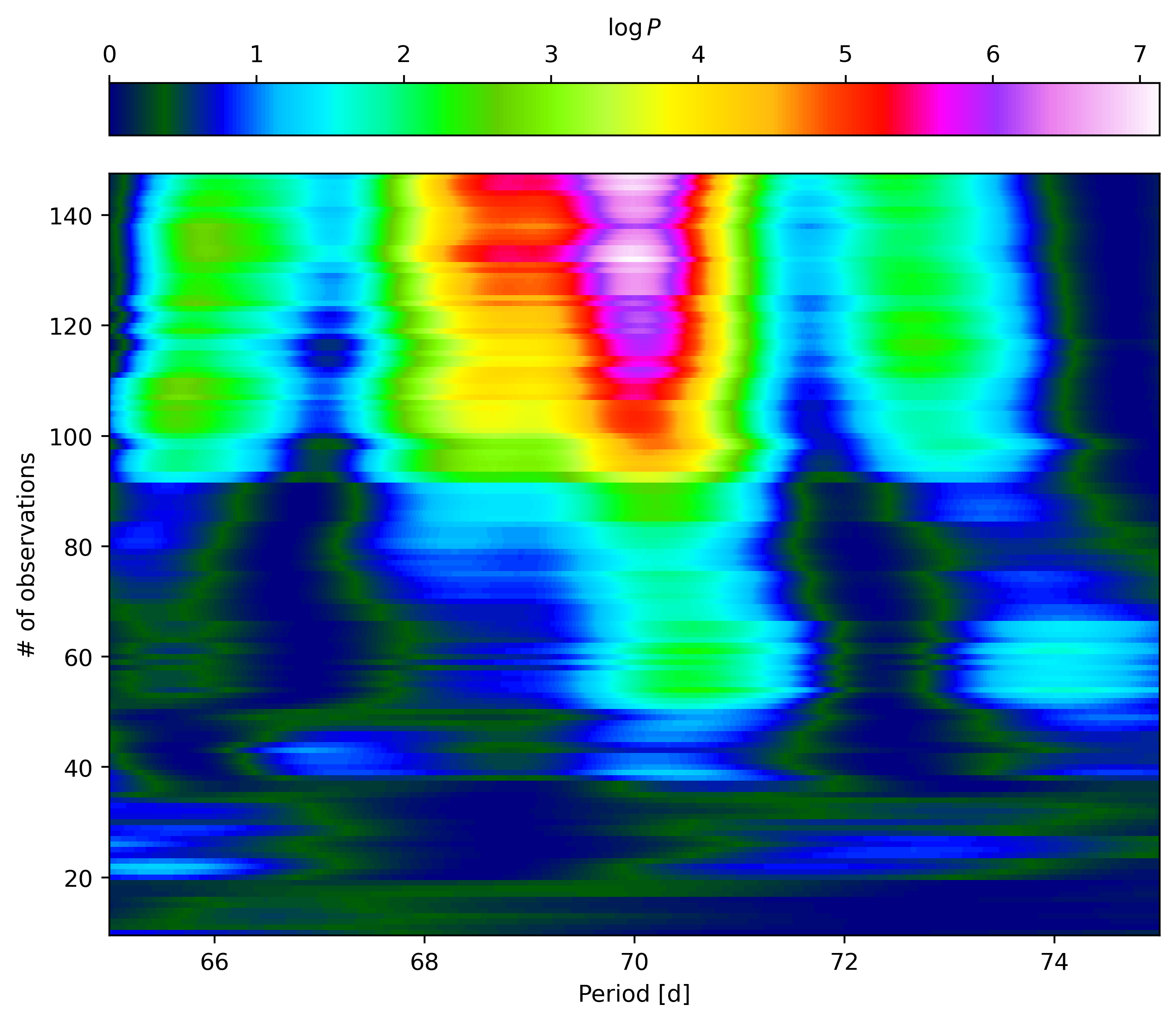} 
    \caption{Stacked Bayesian GLS periodograms of the signals detected in the RVs. \textit{Left panel:} Zoom in on the long-period planetary signal at 1218.5\,d. \textit{Middle panel:} Zoom in on the 2.27\,d period signal. \textit{Right panel:} Zoom in on the 69.94\,d variable signal, which appears in the RV residuals after the 2-planet model is subtracted.}
    \label{fig:sBGLS_J06024}
\end{figure*}

\begin{figure*}[h!]
    \centering
    \includegraphics[width=\textwidth]{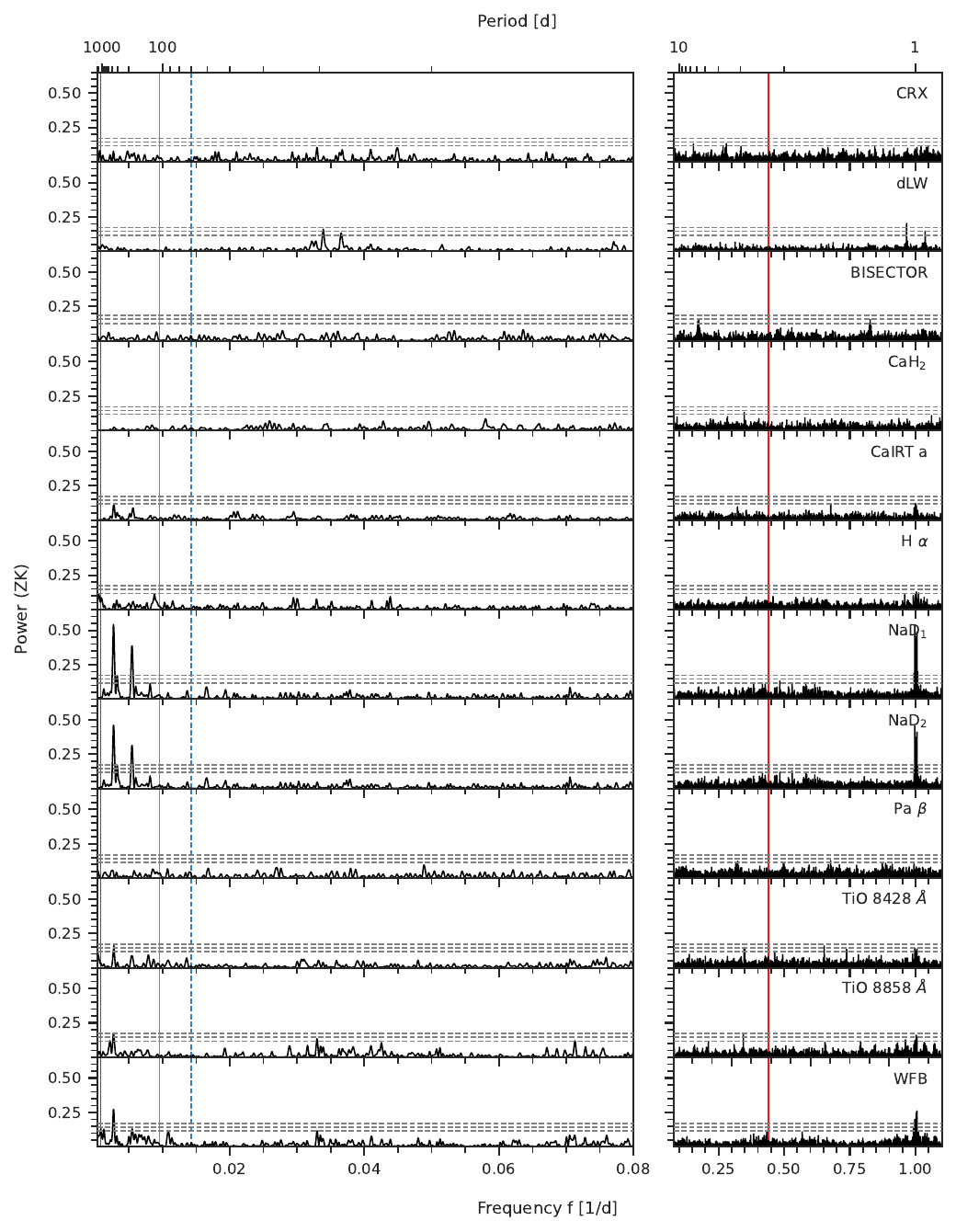}
    \caption{Same as Fig.~\ref{fig:activity_GLS_J01048-181_onecol} but for the activity indicators with significant signals for \StarJTwo{}. The periods 2.27\,d and 1218.5\,d of the planet signals are highlighted by the red solid lines, and the signal found in the residuals at \SI{69.94}{\day} is indicated by the blue dashed line, respectively. The rotation period of \SI{105}{\day} determined by \cite{DiezAlonso2019} is marked by the green solid line.}
    \label{fig:activity_GLS_J06024}
\end{figure*}

\clearpage
\onecolumn
\section{Model priors}

\begin{table*}[h!]
\caption{Priors used for the final RV fits presented in \autoref{sec:meth_and_res}.}
\label{tab:planetparams_priors}
\centering
\begin{tabular}{lcccccp{2cm}}
\hline \hline
\noalign{\smallskip}
Parameter & G~268--110\,b & G~261--6\,b & G~192--15\,b & G~192--15\,c & Unit & Description \\
\noalign{\smallskip}
        \hline
        \noalign{\smallskip}
        \multicolumn{7}{c}{\textit{Planet parameters}} \\
        \noalign{\smallskip}
$P$ & $\mathcal{U}(1.432, 1.4335)$ & $\mathcal{U}(5.2, 5.6)$ & $\mathcal{U}(2.272,2.29)$ & $\mathcal{U}(1000,1500)$ & d & Planetary period \\
$t_{0}$ & $\mathcal{U}(2457613,2457614)$ & $\mathcal{U}(2459342,2459348)$ & $\mathcal{U}(2457850,2457853)$ & $\mathcal{U}(2457850,2459850)$ & BJD & Time of periastron passage \\
$K$ & $\mathcal{U}(0, 10)$ & $\mathcal{U}(0, 15)$ & $\mathcal{U}(0, 5)$ & $\mathcal{U}(2, 10)$ & $\mathrm{m\,s^{-1}}$ & RV semi amplitude \\
$\sqrt{e_\text{b}}\sin \omega_\text{b}$ & fixed(0) & fixed(0) & fixed(0) & $\mathcal{U}(-1, 1)$ & -- &  Parameterization for $e$ and $\omega$\\
$\sqrt{e_\text{b}}\cos \omega_\text{b}$ & fixed(0) & fixed(0) & fixed(0) & $\mathcal{U}(-1, 1)$ &  -- &  Parameterization for $e$ and $\omega$\\

\noalign{\smallskip}
\multicolumn{7}{c}{\textit{Instrument parameters}} \\
\noalign{\smallskip}
$\gamma$        &  $\mathcal{U}(-10, 10)$  & $\mathcal{U}(-10, 10)$           & \multicolumn{2}{c}{$\mathcal{U}(-10, 10)$}        & \si{\meter\per\second} & RV zero point    \\
$\sigma$ &   $\mathcal{U}(0, 10)$     & $\mathcal{U}(0, 10)$ &  \multicolumn{2}{c}{$\mathcal{U}(0, 20)$}& \si{\meter\per\second} & RV jitter added in quadrature \\
\noalign{\smallskip}
\hline
\end{tabular}
\tablefoot{The prior label $\mathcal{U}$ represents uniform distributions.}
\end{table*}

\begin{table*}[h!]
    \centering
    \caption{Default priors used for the dSHO-GP kernel in the fits to photometric data in \autoref{app:rot_period}.}
    \label{tab:priors_GPs}
    \begin{tabularx}{\hsize}{l c c X}
        \hline
        \hline
        \noalign{\smallskip}
        Parameter                                       & Prior                               & Unit                   & Description                                                     \\
        \noalign{\smallskip}
        \hline
        \noalign{\smallskip}
        $P_\text{GP}$                          & $\mathcal{U}(90,110)$, $\mathcal{U}(110,130)$       & d                      & Period    \\
        $\sigma_\text{GP, inst.}$ & $\mathcal{J}(1,\num{1e6})$            & \si{ppm} & Standard deviation of the GP for each data set separately                                             \\
        $f_\text{GP}$                               & $\mathcal{U}(0, 1)$             & ...                  & Fractional amplitude of secondary mode                        \\
        $Q_{0, \text{GP}}$                          & $\mathcal{J}(0.1, \num{1e5})$        & ...                  & Quality factor of secondary mode                              \\
        $dQ_\text{GP}$                              & $\mathcal{J}(0.1, \num{1e5})$        & ...                  & Difference in quality factor between primary and secondary mode \\
        \noalign{\smallskip}
        \hline
    \end{tabularx}
    \tablefoot{The prior labels $\mathcal{U}$ and $\mathcal{J}$ represent uniform and log-uniform distributions, respectively.}
\end{table*}
\end{appendix}

\end{document}